\documentclass[twocolumn]{aastex631}
\pdfoutput=1 
\usepackage[utf8]{inputenc}
\usepackage{amsmath}
\usepackage{bm}
\usepackage{hyperref}

\shorttitle{Distances to Nine Nearby Galaxies using Carbon Stars}
\shortauthors{Zgirski et al.}
\begin{document}

\title{The Araucaria Project. Distances to Nine Galaxies Based on a Statistical Analysis of their Carbon Stars (JAGB Method).}

\author[0000-0003-1515-6107]{Bartłomiej Zgirski}
\affiliation{Nicolaus Copernicus Astronomical Center, Polish Academy of Sciences, Bartycka 18, 00-716, Warszawa, Poland}
\email{bzgirski@camk.edu.pl}

\author{Grzegorz Pietrzyński}
\affiliation{Nicolaus Copernicus Astronomical Center, Polish Academy of Sciences, Bartycka 18, 00-716, Warszawa, Poland}
\affiliation{Universidad de Concepción, Departamento de Astronomia, Casilla 160-C, Concepción, Chile}

\author{Wolfgang Gieren}
\affil{Universidad de Concepción, Departamento de Astronomia, Casilla 160-C, Concepción, Chile} 

\author{Marek Górski}
\affil{Nicolaus Copernicus Astronomical Center, Polish Academy of Sciences, Bartycka 18, 00-716, Warszawa, Poland} 

\author{Piotr Wielgórski}
\affil{Nicolaus Copernicus Astronomical Center, Polish Academy of Sciences, Bartycka 18, 00-716, Warszawa, Poland}  

\author{Paulina Karczmarek}
\affil{Universidad de Concepción, Departamento de Astronomia, Casilla 160-C, Concepción, Chile} 

\author{Fabio Bresolin}
\affil{Institute for Astronomy, 2680 Woodlawn Drive, Honolulu, HI 96822, USA}

\author{Pierre Kervella}
\affil{LESIA, Observatoire de Paris, Université PSL, CNRS, Sorbonne Université, Université de Paris, 5 place Jules Janssen, 92195 Meudon, France} 

\author{Rolf-Peter Kudritzki}
\affil{Institute for Astronomy, 2680 Woodlawn Drive, Honolulu, HI 96822, USA}
\affil{LMU München, Universitätssternwarte, Scheinerstr. 1, 81679 München, Germany}

\author{Jesper Storm}
\affil{Leibniz-Institut für Astrophysik Potsdam (AIP), An der Sternwarte 16, 14482 Potsdam, Germany}

\author{Dariusz Graczyk}
\affil{Nicolaus Copernicus Astronomical Center, Polish Academy of Sciences, Rabiańska 8, 87-100, Toruń, Poland}

\author{Gergely Hajdu}
\affil{Nicolaus Copernicus Astronomical Center, Polish Academy of Sciences, Bartycka 18, 00-716, Warszawa, Poland}

\author{Weronika Narloch}
\affil{Universidad de Concepción, Departamento de Astronomia, Casilla 160-C, Concepción, Chile}

\author{Bogumił Pilecki}
\affil{Nicolaus Copernicus Astronomical Center, Polish Academy of Sciences, Bartycka 18, 00-716, Warszawa, Poland}

\author{Ksenia Suchomska}
\affil{Nicolaus Copernicus Astronomical Center, Polish Academy of Sciences, Bartycka 18, 00-716, Warszawa, Poland}

\author{Mónica Taormina}
\affil{Nicolaus Copernicus Astronomical Center, Polish Academy of Sciences, Bartycka 18, 00-716, Warszawa, Poland}

\begin{abstract}
Our work presents an independent calibration of the J-region Asymptotic Giant Branch (JAGB) method using Infrared Survey Facility (IRSF) photometric data and a custom luminosity function profile to determine JAGB mean magnitudes for nine galaxies. We determine a mean absolute magnitude of carbon stars of $M_{LMC}=-6.212 \pm 0.010$ (stat.) $\pm 0.030$ (syst.) mag.
We then use near-infrared photometry of a number of nearby galaxies, originally obtained by our group to determine their distances from Cepheids using the Leavitt law, in order to independently determine their distances with the JAGB method. We compare the JAGB distances obtained in this work with the Cepheid distances resulting from the same photometry and find very good agreement between the results from the two methods. The mean difference is 0.01 mag with an rms scatter of 0.06 mag after taking into account seven out of the eight analyzed galaxies that had their distances determined using Cepheids. The very accurate distance to the Small Magellanic Cloud (SMC) based on detached eclipsing binaries \citep{SMC-DEB} is also in very good agreement with the distance obtained from carbon stars.
\end{abstract}

\keywords{distance scale --- infrared: stars --- stars: carbon stars --- JAGB method --- galaxies -- galaxies: distances and redshifts}

\section{Introduction}

Although carbon stars have been discovered already in the mid 1800s, their use in distance determinations dates back to the 1980s with NGC 300 being the first object that had its distance determined using them by \citet{RICHER}.
The method has very recently experienced a revival with the papers of \citet{MADORE}, \citet{FREEDMAN}, \citet{RIPOCHE}, \citet{PARADA}, and \citet{LEE}.

The Araucaria Project focuses on an accurate calibration of the cosmic distance scale which is a fundamental issue for many branches in astronomy. In particular, it is crucial for the determination of the present rate of expansion of the Universe, the Hubble constant, which requires an accurate determination of the distances to supernovae host galaxies. The use of diverse distance determination methods allows for better calibrations of the distance scale and the detection of the different systematic uncertainties affecting any of these specific methods. Over the past two decades, our project has produced, among other results, an accurate determination of distances to nearby galaxies using multi-band Cepheid period-luminosity relations. The near-infrared photometric data obtained for these studies (\citealt{N300-CEP}, \citealt{N6822-CEP}, \citealt{N3109-CEP}, \citealt{WLM-CEP}, \citealt{N247-CEP}, \citealt{M33-CEP}, \citealt{N7793-CEP}) are now being utilized for the purpose of calibrating and applying a method that is new to our project, and which relies on the J-region Asymptotic Giant Branch stars (JAGB).

\citet{CMD-POP} identified different stellar populations in the Large Magellanic Cloud (LMC) that occupy different areas in color-magnitude diagrams (CMDs). Among them, the Region J (carbon) stars play a pivotal role in our work. In these thermally pulsating AGB stars the convective envelope brings carbon-enriched material from the helium-burning shell to the surface during thermal pulses (the 3rd and the following dredge-ups). Carbon stars are redder than their oxygen-rich counterparts which they evolved from - the increased C/O ratio changes the molecular opacities of their photospheres which decreases the effective temperatures of such stars \citep{Marigo}, and as a consequence they occupy a redder area in the CMD than the O-rich AGB stars. On the other hand, their magnitudes are confined within a relatively narrow interval, as only stars having masses from a specific range may develop carbon-rich photospheres. Too massive AGB stars experience hot bottom burning by converting carbon into nitrogen at the base of their convective envelopes - the exact threshold stellar mass that allows for such a process depends strongly on metallicity according to different models (e.g. \citealt{Karakas}, \citealt{Ventura}). Stars with low masses of $M<1.3M_\sun$ don't have their convective zones developed enough to bring carbon to the surface (\citealt{G&M}, \citealt{Iben}, \citealt{Sackmann}). These facts suggest that carbon star luminosity functions might serve as useful standard candles. Important advantages of the use of carbon stars for distance determinations are their high luminosities in the infrared, and the opportunity to determine the distance of a given galaxy from just one epoch of observation. In the near-infrared $J-$ band, JAGB stars populate an area of approximately constant magnitude (with a standard deviation spread $\sigma$).

The aim of this work is to use archival data of the Araucaria Project for a number of nearby galaxies for the purpose of testing the potential of the JAGB method, by checking its precision and accuracy through a comparison of the resulting JAGB distances with the Cepheid distances obtained from the same data. In our work, we propose an alternative method of determining mean JAGB magnitudes and use them as standard candles. The method relies on fitting of the luminosity function adopting a superposition of a Gaussian and a quadratic function. We calibrate the method using data from the Large Magellanic Cloud, and determine the distances to nine other nearby galaxies. Our method yields distance results in excellent agreement with those obtained by using the Leavitt law for classical Cepheids. The stability of the results with respect to different sample-selection box positions and sizes is improved by our profile fitting method when comparing the ordinary median or average of stellar magnitudes in a box.

\section{Definition and calibration of the method}
\begin{figure}[h]
\plotone{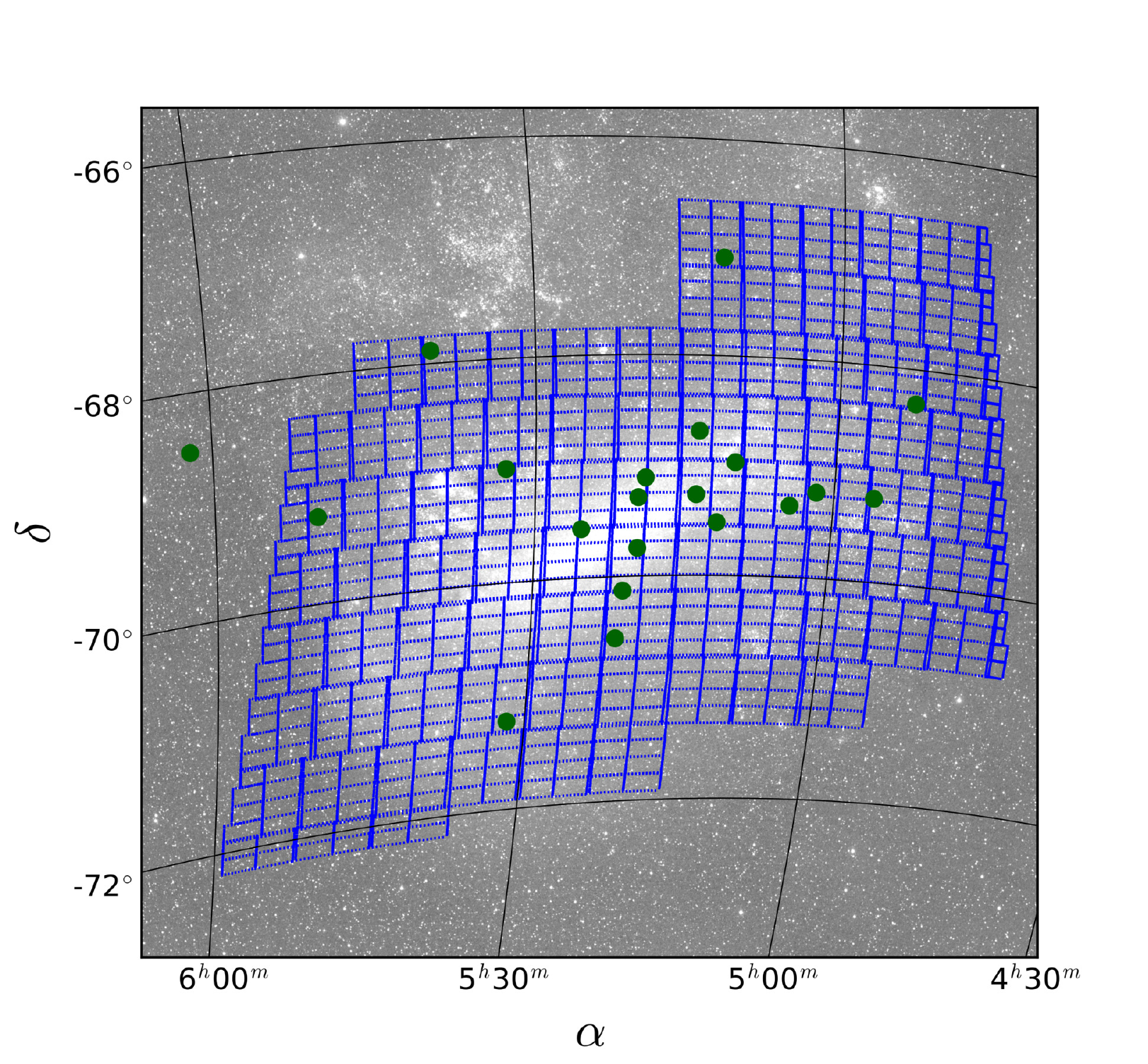}
\caption{Location of the detached eclipsing binaries utilized to determine the LMC distance in \citet{LMC-DEB}, together with the IRSF \citep{IRSF-CAT} fields used for the calibration of the average J magnitude of the JAGB population in this galaxy. Background image from The All Sky Automated Survey originates from \citet{UDALSKI}.}
\label{EBinLMC}
\end{figure}

The calibration of the method comes down to the determination of the expected value of absolute magnitudes of JAGB stars. We rely on the distance to the LMC - $\mu_0=18.477 \pm 0.004$ (stat.) $\pm 0.026$ (syst.) mag - accurately determined using detached eclipsing binaries \citep{LMC-DEB}. We assume the JAGB populations of this galaxy to be fiducial for our distance determinations using the method. Near-Infrared $J-$ and $K-$ band photometry of the Magellanic Clouds (MCs) comes from the IRSF Magellanic Clouds Point Source Catalog \citep{IRSF-CAT}. We have converted the photometry from IRSF/SIRIUS into the 2MASS photometric system according to conversion equations given in that work.
IRSF photometric maps of the main body of the LMC correspond to the areas where the detached eclipsing binaries used to determine the LMC distance are located. Figure \ref{EBinLMC} depicts the location of these binary systems in the area where the the LMC JAGB sample comes from.
We have dereddened each photometric field separately, assuming the corresponding reddening from the \citet{MC-EXT} Magellanic Clouds extinction maps. We used the ratios of total-to-selective extinctions of \citet{Cardelli} in order to convert $E(B-V)$ selective extinctions given in these works into $A_J$ and $A_K$ total extinctions. From this procedure we obtained the reddening-free color-magnitude diagram for the LMC.
It is worth noticing here that in the calibration by \citet{MADORE}, the authors assume $A_{J, LMC}=0.053$ mag and $A_{J, SMC}=0.026$ mag based on \citet{Schlafly}. The authors note that these values include only foreground extinction, neglecting the internal component of extinction within the MCs. Their approach is however consistent, as they stick to the same procedure in the determinations of the extinctions to their target galaxies.
\citet{RIPOCHE} adopt the extinction from \citet{MC-EXT} by taking one (average) value of the reddening for the whole LMC.

\begin{figure}[h]
\plotone{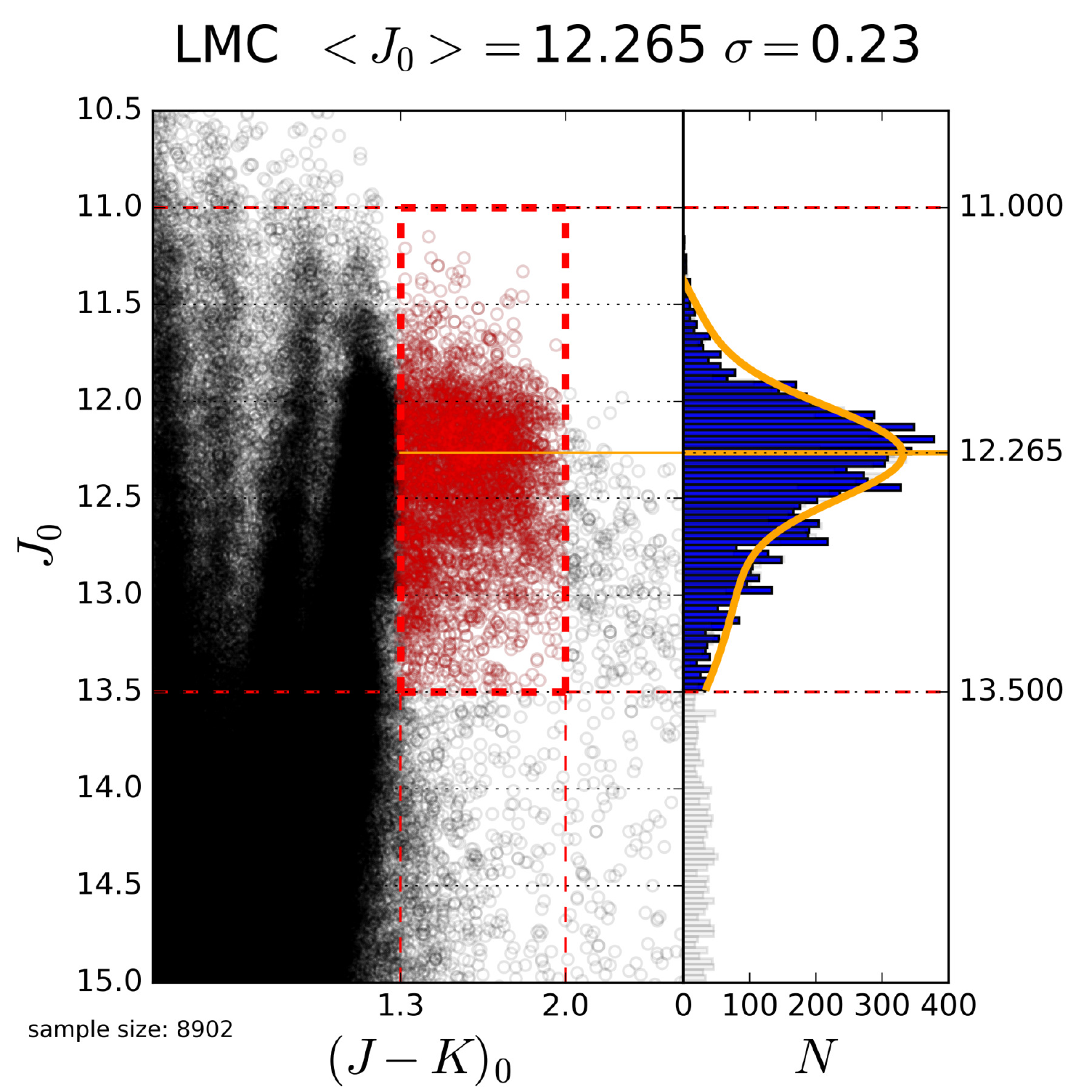}
\caption{Determination of the expected value of the $J$ magnitude for the population of carbon stars in the LMC for $1.30<(J-K)<2.0$ mag.}
\label{LMC_deter1.3}
\end{figure}

\begin{figure}[h]
\plotone{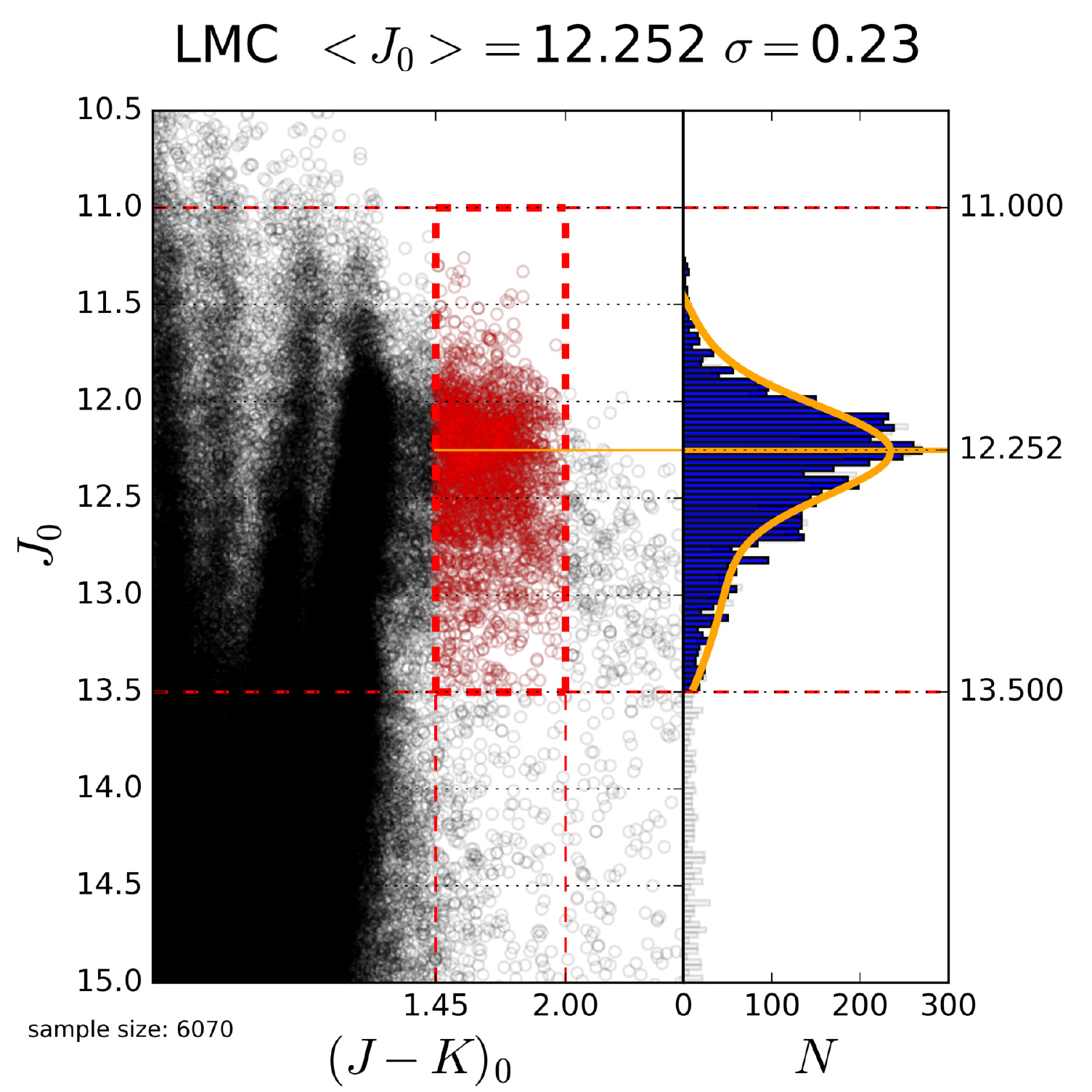}
\caption{As in Figure \ref{LMC_deter1.3}, but for $1.45<(J-K)<2.0$ mag. This color interval will be used for some heavily contaminated populations, however the calibration of the method relies on the larger color interval depicted in Figure \ref{LMC_deter1.3}.}
\label{LMC_deter1.45}
\end{figure}

We have established luminosity functions of JAGB stars for the LMC by binning the stellar $J$ magnitudes into histograms. In our work, the mean dereddened, apparent $<J_0>$ magnitude was determined as the mean value of the Gaussian component of the fitted profile. The latter is a superposition of a Gaussian and a quadratic function:

\small
\begin{equation}
P(x)=\frac{N}{\sigma \sqrt{2\pi} } \exp{\bigg[\frac{-(x-m)^2}{2\sigma^2}\bigg]}+a(x-m)^2+b(x-m) + c 
\end{equation}
\normalsize

where $m$ is our estimate of the mean apparent magnitude of a given JAGB population. The quadratic component takes into account foreground stars and background galaxies that are not members of the population of carbon stars and contaminate our samples.
We used this model before for the purpose of determining mean colors of red clump stars (e.g. \citealt{MC-EXT}). The six-parameter fits of the profile were performed using the \textit{curve\_fit} function of SciPy \citep{SCIPY}.
We have defined the area of the CMD populated by JAGB stars to be limited to the color interval $1.30<(J-K)<2.0$ mag. However, for populations visibly contaminated from the bluer side, we adopted $1.45<(J-K)<2.0$ mag. Figures \ref{LMC_deter1.3} and \ref{LMC_deter1.45} show CMDs and luminosity functions for stars in the LMC for the two color intervals mentioned above - as can be appreciated, the two determinations agree with each other at the scale of about 0.01 mag. We base our final calibration on the broader color interval $1.30<(J-K)<2.0$ mag. 

We adopted a uniform width of the sample selection box of $\Delta J_0=2.5$ mag for all galaxies we analyzed. The box width along the $J-$ axis requires discussion (see further discussion section). Finally, the expected mean value of the carbon star $J-$ magnitude that will serve as our fiducial is $<J_0>_{LMC}=12.265$ mag. For comparison, the average and median values are $12.40$ mag, and $12.35$ mag, respectively. The standard deviation ($\sigma_{LMC}=0.23$ mag) of the Gaussian component of the luminosity function includes factors such as the intrinsic spread in the absolute magnitudes of the population of JAGB stars, photometric errors, differential extinction, and differences in geometric depth of stars in the system \citep{Weinberg}. We have also determined the rms scatter of $<J_0>_{LMC}$ using a bootstrap method and obtain a value of $0.01$ mag, which we adopt as the statistical uncertainty of our determination.
By subtracting the distance modulus of the LMC, we obtain the true, dereddened absolute $J$ mean magnitude of the JAGB population in the LMC: $M_{LMC}=-6.212 \pm 0.010$ (stat.) $\pm 0.030$ (syst.) mag. We have decided to rely only on this LMC-based calibration since firstly the LMC distance is more accurately measured than the SMC distance due to the more complicated geometrical structure of the SMC (e.g. \citealt{SMC-DEB}), and secondly the JAGB sample in the LMC is larger than that of the SMC. 
It is worth mentioning here that we have tried different statistics to estimate the expected value of $<J_0>$. While the bootstrap method yielded a smaller uncertainty on the mean value than for the $m$ coefficient of the fitted profile in a given box, the stability of the results for different box sizes and centers was lower in the case of adopting the median and mean (see discussion section).
For comparison, \citet{MADORE} obtained an average value of $M=-6.20$ mag from $M_{LMC}=-6.22 \pm 0.01$(stat.)$\pm 0.04$(syst.) mag, $M_{SMC}=-6.18 \pm 0.01$ (stat.)$\pm 0.05$(syst.) mag by taking mean values of magnitudes from JAGB samples, while \citet{RIPOCHE} and \citet{PARADA} have calibrated and used $M_{LMC}=-6.28 \pm 0.004$(stat.) mag, $M_{SMC}=-6.16 \pm 0.015$(stat.) mag by taking median values of these magnitudes. 

\section{Distance determinations to selected galaxies}

Hereafter, all described photometric data, except for the SMC, have been previously published by us as part of our work on distance determinations to nearby galaxies using period-luminosity relations for Cepheids. In order to obtain values compatible with the mean value of the absolute magnitude of the JAGB reported above, we have converted the whole photometry from UKIRT to the 2MASS system using conversion formulae from \citet{Carpenter}. The dereddening of the CMDs for our sample of galaxies is based on the extinction values taken from our previous papers on Cepheid distances - the reported $E(B-V)$ values. In the case of the SMC, we took data from the same database as for the LMC, and we dereddened them in the same way as described in the previous section. 
Cepheid distances based on the Leavitt Law have been determined by our group for all of the following galaxies during the past two decades. They are our main reference for assesing the accuracy of the distances derived with the JAGB method. However, we have refined the Cepheid distances reported in the original papers by tying them to the most recent result for the LMC distance of $18.477$ mag \citep{LMC-DEB} - which has been used for our zero-point calibration. Almost all galaxies analyzed in our previous work had their Cepheid distances tied to an LMC distance modulus of $18.5$ mag. The only exception is NGC 7793 whose reported Cepheid distance was tied to an LMC distance of $18.493$ mag \citep{LMC-2pDEB}. Essentially, the distance moduli for the analyzed galaxies were improved by shifting their values by $0.016$ mag in the case of NGC 7793, and by $0.023$ mag for the remainder of the galaxies.
Just like in the case of the JAGB calibration, we took the rms scatter resulting from bootstrapping of the expected value of the Gaussian component of the fitted profiles ($m$ from Eq. 1) as statistical uncertainties of our determinations. Tables \ref{j0}-\ref{dist_cep_comp} summarize the results reported in this section and their comparison with Cepheid distances obtained from the same data. Table \ref{J_app_moduli} contains determinations of JAGB distance moduli that were obtained using observed, apparent magnitudes with no dereddening procedure applied to the CMDs, while Tables \ref{j0}, \ref{dist_cep_comp} present determinations deriving from dereddened data.

\subsection{SMC}
\begin{figure}[h!]
\plotone{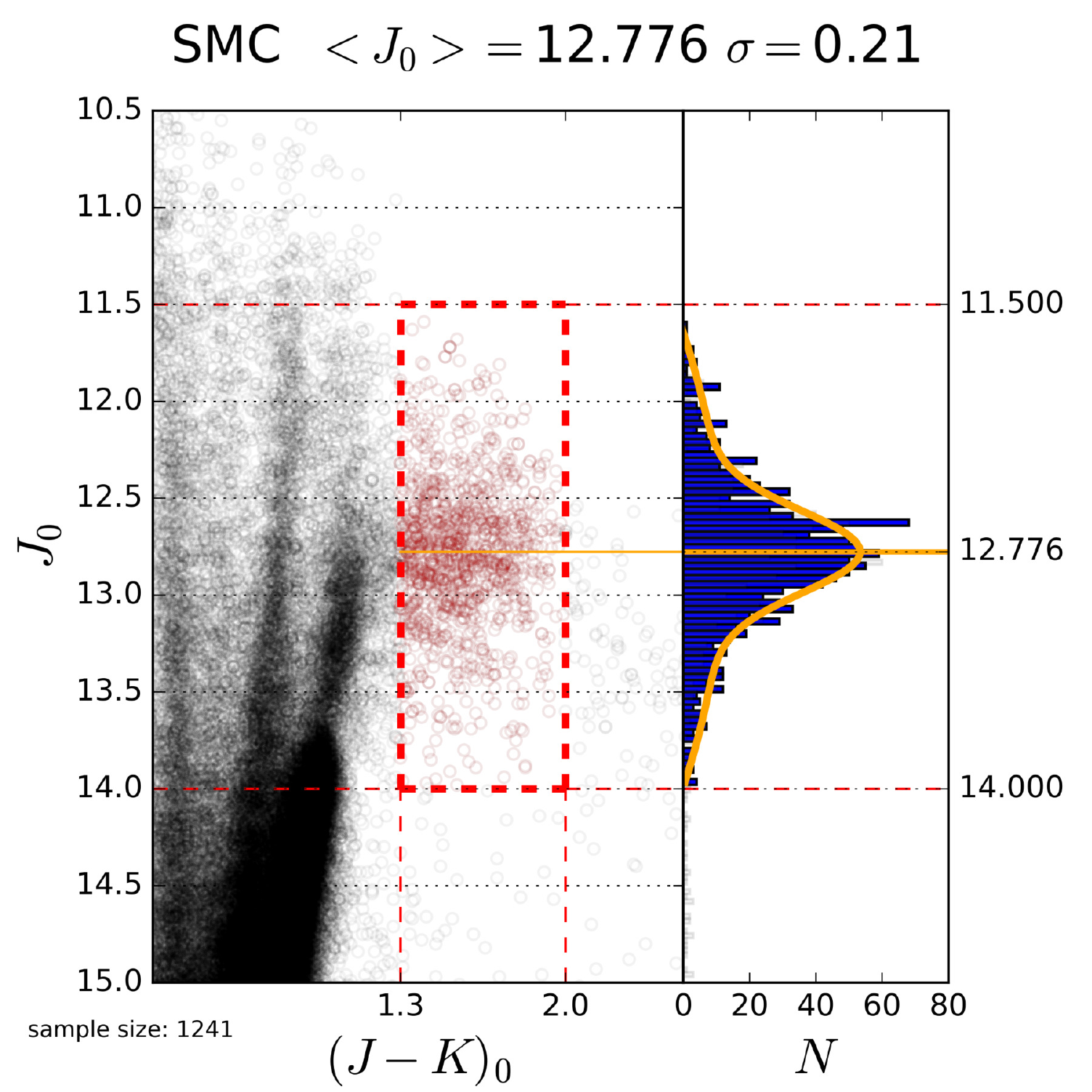}
\caption{Determination of the expected value of the $J$ magnitude for the population of carbon stars in the SMC.}
\label{SMC_deter}
\end{figure}
Figure \ref{SMC_deter} shows the CMD and the luminosity function and its fit for the case of the SMC. 
We notice here that the luminosity function is far more symmetrical than in the case of the LMC. We obtain $<J_0>_{SMC}=12.776 \pm 0.012$ mag. The bootstrap method yields a scatter of $0.012$ mag. With the absolute JAGB magnitude as calibrated above we obtain a distance modulus $\mu_{SMC}=18.988 \pm 0.012$ (stat.) $\pm 0.047$ (syst.) mag for the SMC which is in very good agreement with the distance of $18.977 \pm 0.044$ mag for the SMC obtained by \citet{SMC-DEB} from eclipsing binaries. We also derive the absolute JAGB magnitude of the SMC sample adopting the \citeauthor{SMC-DEB} distance: $M_{SMC}=-6.201 \pm 0.012$ (stat.) $\pm 0.044$ (syst.) mag. This value is in excellent agreement with the JAGB absolute magnitude we obtain for the LMC - contrary to larger discrepancies between the JAGB absolute magnitudes of the MCs reported by others (see the end of the previous section). 

\subsection{NGC 3109}

Using previously published data on NGC 3109 obtained with the ISAAC near-infrared camera installed at the ESO VLT \citep{N3109-CEP}, and the selective extinction to the galaxy $E(B-V)_{N3109}=0.087$ mag provided there, we have fitted the luminosity function to the JAGB sample and obtain $<J_0>_{N3109}=19.305 \pm 0.046$ mag. The corresponding distance modulus is: $\mu_{N3109}=25.517 \pm 0.046$ (stat.) mag. Cepheid period-luminosity relations in the $VIJK$ bands published by \citeauthor{N3109-CEP} yield a distance modulus $\mu_{N3109-CEP}=25.548 \pm 0.024$ (stat.) mag after adoption of the corrected distance to the LMC. These two values are again in excellent agreement, and are also in good agreement with the \citet{FREEDMAN} JAGB determination of $\mu_{N3109-FM}=25.56 \pm 0.05$ mag.

\subsection{NGC 247}
\begin{figure}[h]
\plotone{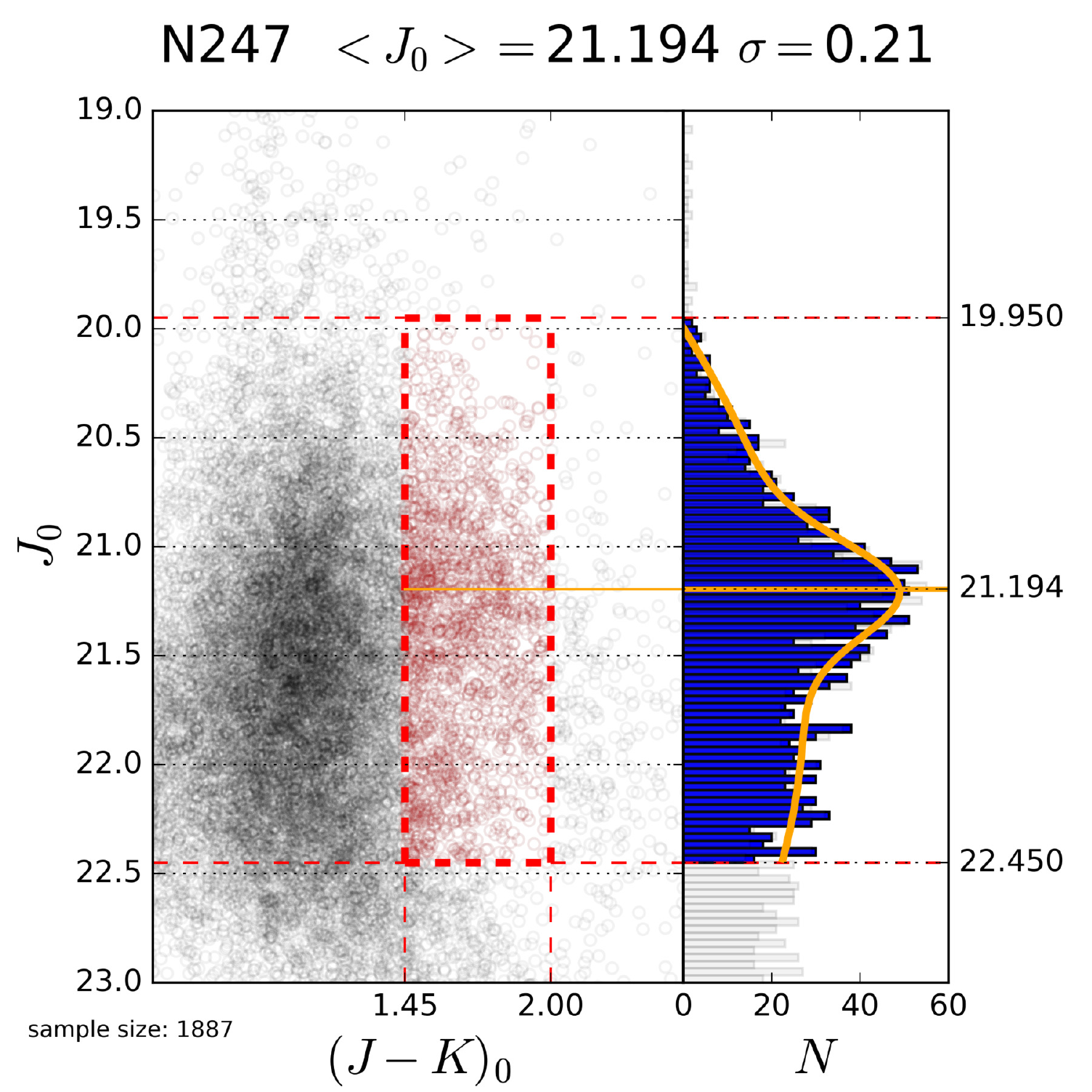}
\caption{Determination of the expected value of the $J$ magnitude for the population of carbon stars in NGC 247.}
\label{N247_deter}
\end{figure}

This Sculptor Group galaxy has ISAAC near-infrared photometry published by \citet{N247-CEP} who determine a mean reddening $E(B-V)_{N247}=0.18$ mag. Figure \ref{N247_deter} depicts the dereddened CMD, together with the area selected to contain carbon stars and the result of the luminosity function fit. 
The adopted color boundaries are different this time - $(J-K)_0 \in (1.45, 2.0)$ mag, as we wanted to exclude contamination caused by O-rich AGB stars. We obtain $<J_0>_{N247}=21.194$ mag. Bootstrap yields an uncertainty of this value of $0.023$ mag. The observed JAGB magnitude yields a distance modulus $\mu_{N247}=27.406 \pm 0.023$ (stat.) mag. The distance to NGC 247 derived by \citet{N247-CEP} using Cepheid variables yielded $\mu_{N247-CEP}=27.621 \pm 0.036$ (stat.) mag after correcting for the improved distance to the LMC. In this case, the distances from the JAGB method and Cepheids do not agree with each other, even after taking into account the known systematic errors.

\pagebreak
\subsection{NGC 55}
This galaxy, also a Sculptor Group member, was observed during two nights using the ISAAC camera \citep{N55-CEP}. In that work a mean reddening $E(B-V)_{N55}=0.127$ mag was determined for the Cepheids. As in the case of NGC 247, the lower limiting value for the $(J-K)_0$ color of the carbon star selection was chosen as $1.45$ mag in order to decrease the contamination of our sample. 
The profile fit yields $<J_0>_{N55_1}=20.124$ mag for the first, and $<J_0>_{N55_2}=20.166$ mag for the second night. We decided to take the average of these two determinations and obtain $<J_0>_{N55}=20.145 \pm 0.019$ mag (with the same bootstrap-determined uncertainties for the two nights). This value yields a distance modulus $\mu_{N55}=26.357 \pm 0.019$ (stat.) mag from the JAGB method. \citet{N55-CEP} found $\mu_{N55-CEP}=26.411 \pm 0.037$ (stat.) mag from Cepheids, again after applying the LMC distance correction. These two results are in agreement even within the statistical errors.

\subsection{NGC 6822}
This Local Group galaxy has been observed in the near-infrared by our group using the SOFI camera installed on the New Technology Telescope (NTT) at La Silla, and independently using the PANIC instrument attached to the Magellan-Baade telescope at Las Campanas Observatory, for the purpose of determining period-luminosity relations for its Cepheids \citep{N6822-CEP}. That work yielded a mean extinction $E(B-V)_{N6822}=0.356$ mag for the galaxy. From the SOFI data, we have obtained $<J_0>_{S,N6822}=16.976 \pm 0.041$ mag. From the PANIC data, the fit yields $<J_0>_{P,N6822}=17.085 \pm 0.032$ mag. By taking the mean value, we obtain $<J_0>_{N6822}=17.031 \pm 0.037$ mag, yielding $\mu=23.243 \pm 0.037$ (stat.) mag for the distance modulus of NGC 6822.

\citet{N6822-CEP} obtained, considering the LMC distance correction, a Cepheid-based distance modulus $\mu_{N6822-CEP-G}=23.289 \pm 0.021$ (stat.) mag which agrees well with the JAGB result presented above. \citet{N6822-CEP-Madore} determined $\mu_{N6822-CEP-M}=23.49 \pm 0.03$ (stat.) mag also using Cepheids. \citet{FREEDMAN} determined a JAGB distance to NGC 6822, although due to the large foreground extinction, they included it in an appendix to their work and not in the main part of the paper. The value they obtained, $\mu_{N6822-JAGB-FM}=23.44 \pm 0.02$ (stat.) mag does not agree with our determination. This is most probably caused by and apparent \citep{N6822-CEP-Madore} systematic shift of the zero point of the near-infrared photometry used in \citet{N6822-CEP}. \citet{PARADA} determined an independent JAGB distance modulus $\mu_{N6822-JAGB-P}=23.54 \pm 0.03$ (stat.) mag, and \citet{Whitelock} obtained $\mu_{N6822-MIR-W}=23.56 \pm 0.03$ (stat.) mag using Miras. We note that the zero point of the J- and K- photometry used here doesn't affect the difference between the distances obtained from both methods, JAGB and Cepheids \citet{N6822-CEP}, because in both determinations the same datasets and calibrations were used.

\subsection{NGC 7793}
The photometric data gathered on two nights with the HAWK-I camera installed on the ESO VLT were taken from \citet{N7793-CEP}. We decided to use the data of the night with the better photometric quality. Also, we adopted for our analysis only data coming from just two out of the four chips of the HAWK-I detector in order to exclude high-density areas in the center of the galaxy with considerable blending, and an area that is mostly made up of field stars (see the chart depicting the observed field in the \citeauthor{N7793-CEP} paper). We also changed the lower limiting color of our sample to 1.45. Our data yielded $<J_0>_{N7793}=21.489 \pm 0.027$ mag. Adopting $E(B-V)_{N7793}=0.08$ mag from \citeauthor{N7793-CEP}, we obtained a distance modulus of $mu_{N7793}=27.701 \pm 0.027$ (stat.) mag. The Cepheid distance to NGC 7793 obtained by \citet{N7793-CEP} is, after the marginal LMC distance adjustment to the \citet{LMC-DEB} value, $ \mu_{N7793-CEP}=27.65 \pm 0.04$ (stat.) mag which is in very good agreement with the result we have obtained from the JAGB technique in this paper.

\pagebreak
\subsection{M33}

\begin{figure}[h!]
\plotone{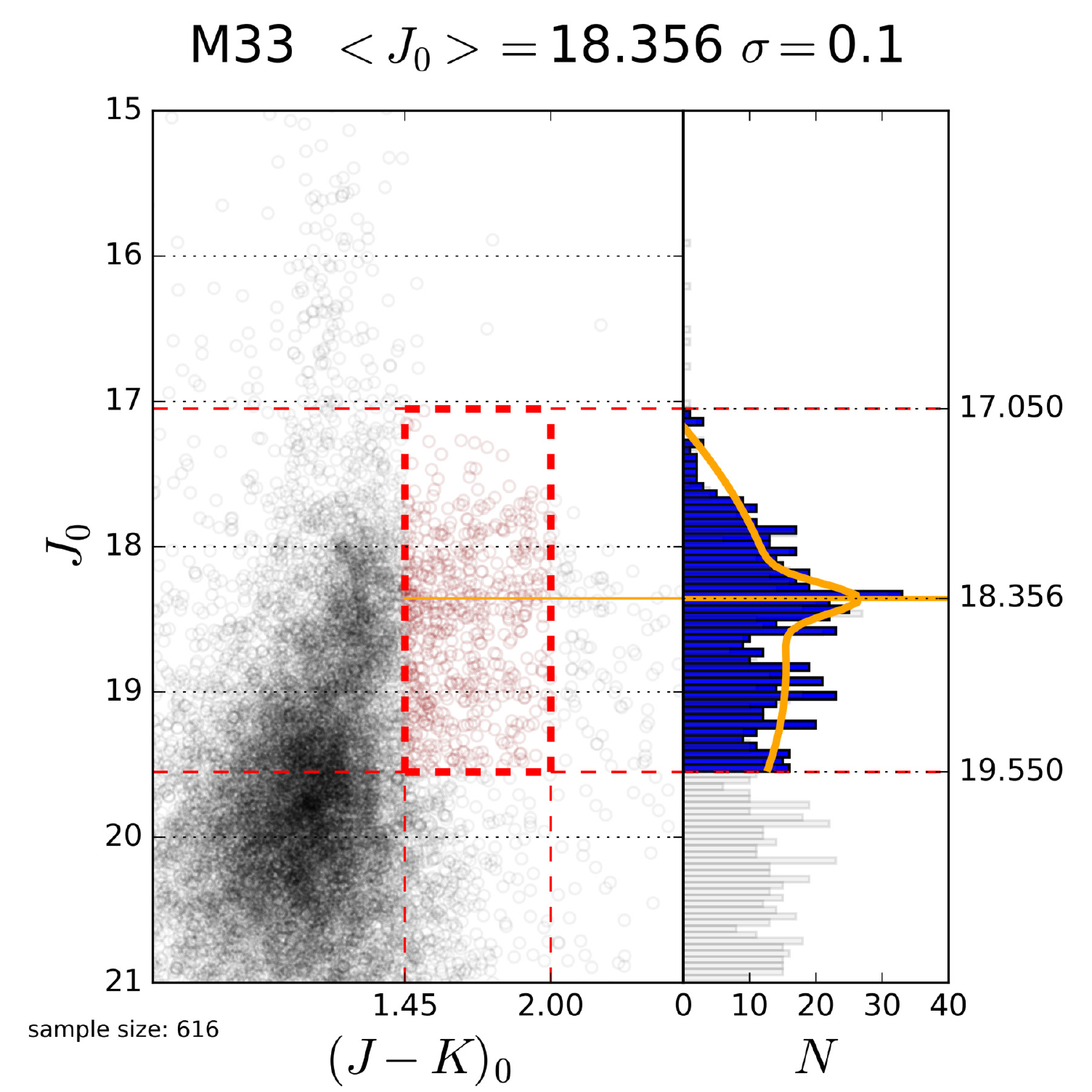}
\caption{Determination of the mean value of the $J$ magnitude for the population of carbon stars in M33.}
\label{M33_deter}
\end{figure}

The data for this galaxy were taken from \citet{M33-CEP}. These were gathered using the HAWK-I camera - we have again chosen two datasets out of four originating from different chips of the camera, excluding the central dense field. \citeauthor{M33-CEP} derived $E(B-V)_{M33}=0.19$ mag as the mean color excess towards the galaxy, which we adopt here.
Figure \ref{M33_deter} shows the CMD and the luminosity function which is characterized by a large amount of contamination and a narrowly peaked maximum, with a small standard deviation $\sigma$. 

\begin{figure}[h!]
\plotone{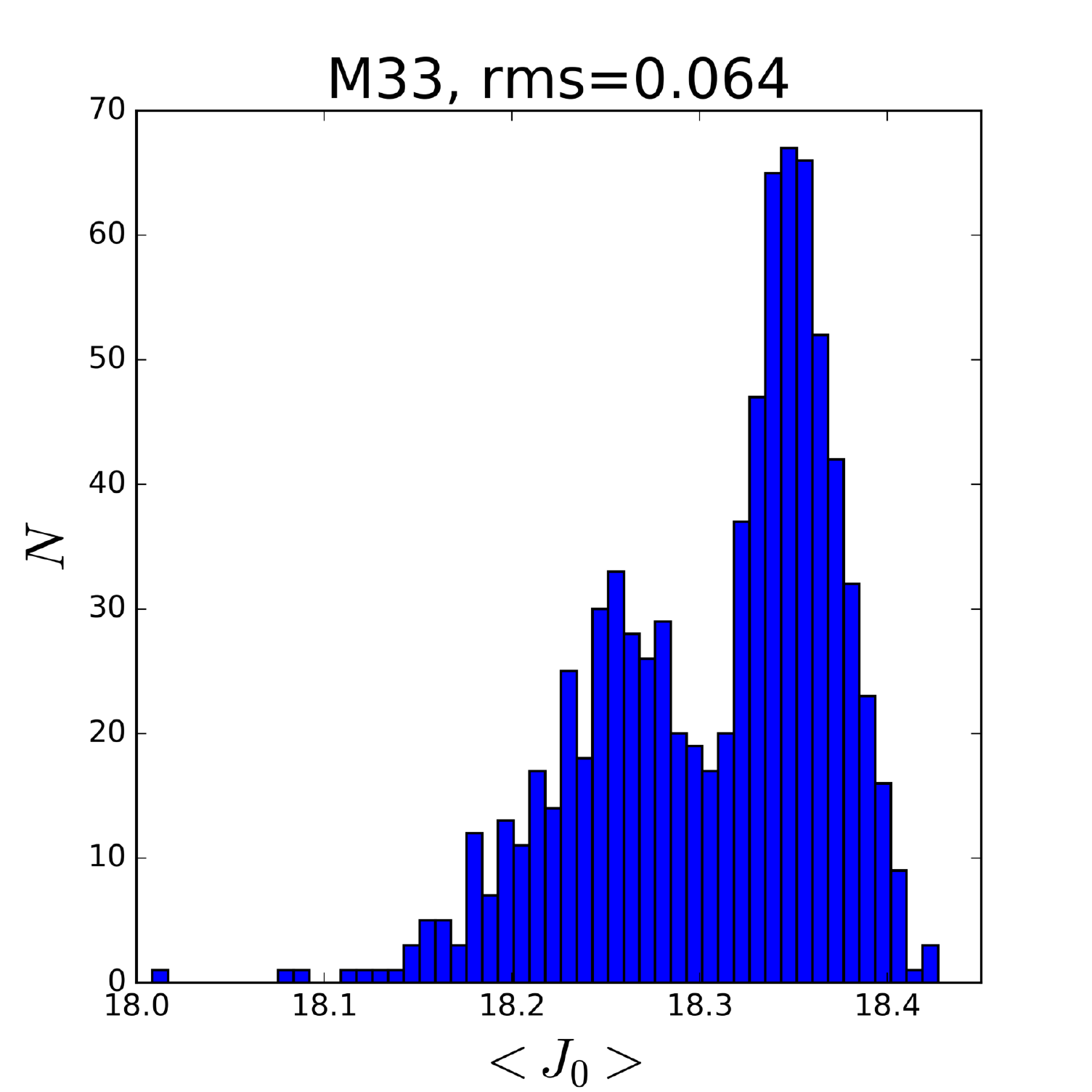}
\caption{Distribution of $\mu$ (the expected value of the Gaussian component) of the fitted luminosity function of carbon stars from M33 modeled using the bootstrap method.}
\label{muM33}
\end{figure}

We have obtained $<J_0>_{M33}=18.356$ mag. Bootstrap simulations yielded an rms scatter of $0.064$ mag. The distribution of expected values extends towards smaller values (usually it resembles a normal distribution). Apparently the highly contaminated sample allows for a degeneration of the fitted expected luminosities of the JAGB - see Fig. \ref{muM33}. Our result lies, however, near the maximum of the distribution of the simulated outcomes. We keep the large uncertainty returned by the bootstrap simulations: $\mu_{M33}=24.568 \pm 0.064$ (stat.) mag. \citet{M33-CEP} determined the Cepheid distance to the galaxy, corrected to the new LMC distance, to be $\mu_{M33-CEP}=24.60 \pm 0.03$ (stat.) mag which is in agreement, within the uncertainties, with the JAGB distance determination in this work.

\subsection{NGC 300}
The data for this galaxy originate from \citet{N300-CEP}, and were gathered using ISAAC infrared camera on the ESO VLT. There are two samples from two separate observing nights containing more than 100 stars. Due to the large contamination for the lower values of the $(J-K)_0$ color, we have selected $1.45$ mag as the lower boundary of the sample-selection box. Bootstrapping yields a $0.06$ mag spread around the mean magnitudes. The results from the data of the two nights are:$<J_0>_{1, N300}=20.296 \pm 0.061$ mag; $<J_0>_{2,N300}=20.217 \pm 0.065$ mag. 
After taking the mean value of the two results and adopting the extinction $E(B-V)_{N300}=0.096$ mag derived in the \citeauthor{N300-CEP} study, we obtain a distance modulus $\mu_{N300}=26.469 \pm 0.063$ (stat.) mag which is in reasonable agreement with the \citet{N300-CEP} value of $\mu_{CEP-N300}=26.344 \pm 0.04$ (stat.) mag derived from Cepheids (again, tied to the most recent LMC distance determination). 

\subsection{Wolf–Lundmark–Melotte (WLM)}
\begin{figure}[h]
\plotone{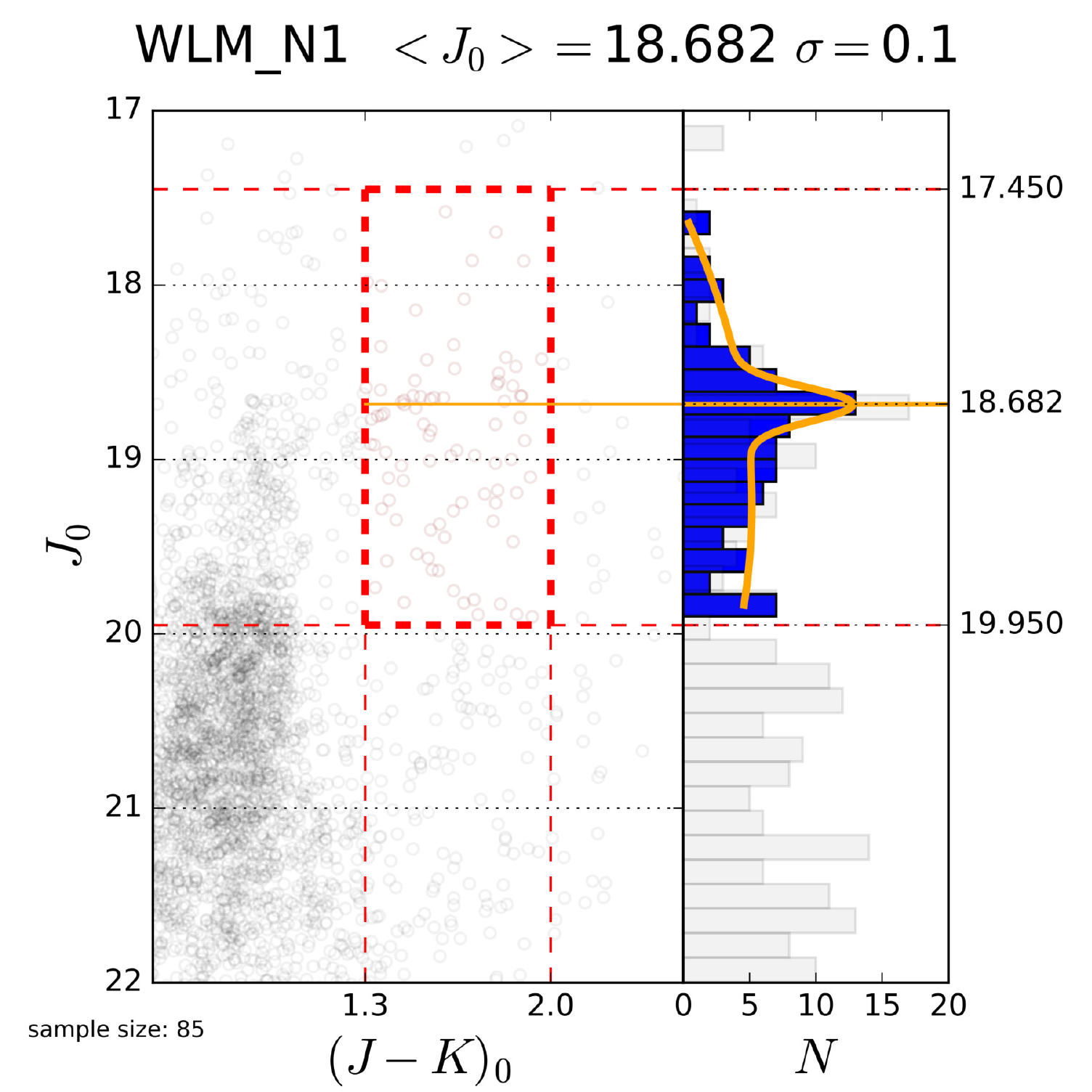}
\caption{Determination of the expected value of the $J$ magnitude for the population of carbon stars of WLM (first night).}
\label{WLM_deter1}
\end{figure}

The WLM data adopted from the \citet{WLM-CEP} study include small samples of JAGB stars. The reddening determined for this galaxy in that paper is $E(B-V)_{WLM}=0.082$ mag. We used data from two consecutive nights that established two different samples. The luminosity function of our sample derived from the first night (Figure \ref{WLM_deter1}) is characterized by a narrow peak which yields a very small standard deviation. While usually this is undesirable and we didn't consider fits with spreads smaller than $0.075$ mag as they often tended to pick up noise or fluctuations, in this case the sample is one of the smallest among those analyzed by us. We decreased the bin density (to 0.7 bins per $0.1$ mag, before we used 1-3 bins per $0.1$ mags, depending on the sample sizes) which yielded virtually the same results in terms of $<J_0>_{1,WLM}$ but increased $\sigma$ which cannot be determined reliably for such a sample. 
This is not the case for the dataset from the second night where the spread is larger and virtually the same for different binning (around $0.2$ mag). As the sample sizes for each night are comparable, we stick to one bin size for both.
Overall, for the first night, we obtained $<J_0>_{1, WLM}=18.682 \pm 0.080$ mag. The data from the second night yield $<J_0>_{2,WLM}=18.802 \pm 0.080$ mag. Large uncertainties remain even for smaller bin sizes in the case of the second night and they underline the uncertainty introduced by small and rather significantly contaminated samples.
By taking the average $<J_0>$ from the two nights and adopting the previously calibrated absolute magnitude of the JAGB, we obtain a distance modulus $\mu_{WLM}=24.954 \pm 0.08$ (stat.) mag. The \citet{WLM-CEP} work results in $\mu_{WLM-CEP}=24.901 \pm 0.042$ (stat., value corrected for the $18.477$ LMC distance modulus) mag and is in good agreement with our present value derived from the JAGB method. \citet{LEE} obtained $\mu_{WLM-P}=24.97 \pm 0.02$ (stat.) $\pm 0.04 $(syst.) mag also from the JAGB technique - their result is in very good agreement with our present determination.

\section{Comparison with Cepheid distances}
\begin{figure}[t]
\plotone{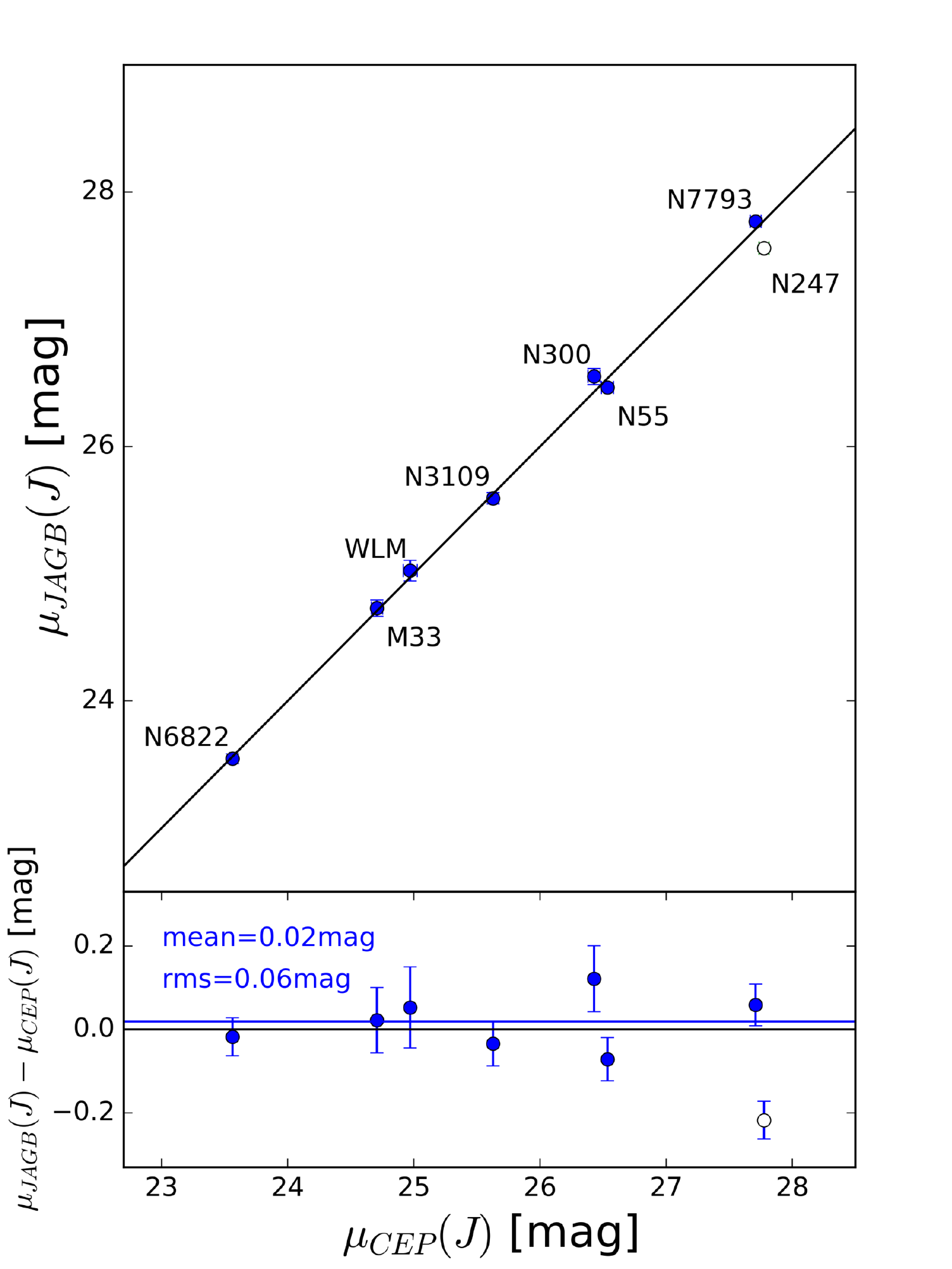}
\caption{Comparison between published Cepheid apparent distance moduli in the $J-$band and JAGB apparent distance moduli. The black line corresponds to the 1-1 relation between Cepheid and JAGB distances and is plotted as reference.}
\label{COMPARISON_MOD}
\end{figure}

\begin{figure}[t]
\plotone{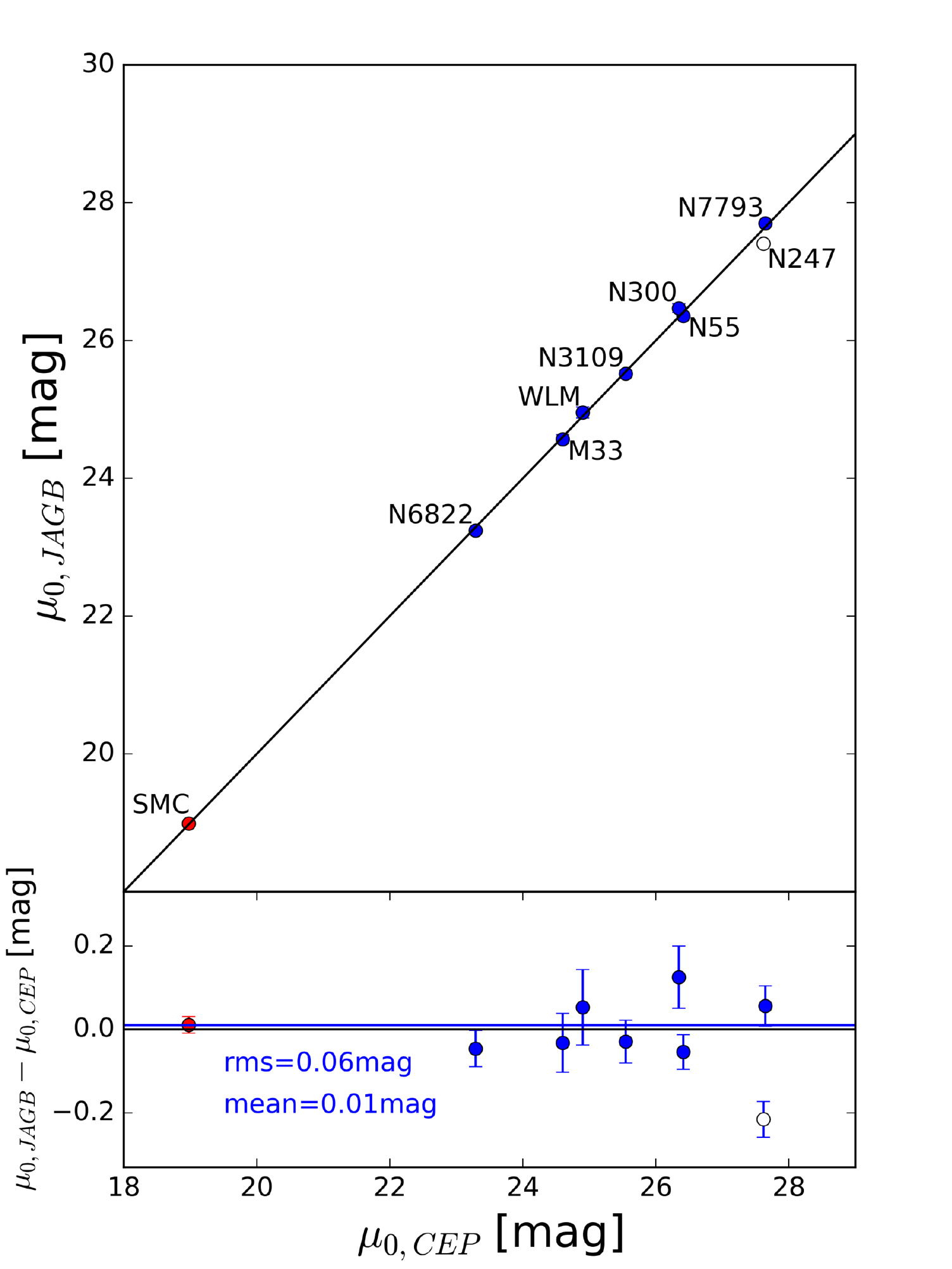}
\caption{Comparison between published Cepheid distances (unreddened distance moduli) and JAGB distances which were obtained in this work. The black line corresponds to the 1-1 relation between Cepheid and JAGB distances and is plotted as reference. The red dot denoting the SMC is an exception, and shows the comparison between the present JAGB distance, and the eclipsing binaries distance of \citet{SMC-DEB}.}
\label{COMPARISON}
\end{figure}

Since the data used in this paper have been previously employed to determine the distances of our sample galaxies with the Cepheid period-luminosity relation, a comparison of our new JAGB distances to the Cepheid distances provides an excellent test of the capabilities of the JAGB method. The comparison of the distance results provided by the two methods is given in Tables \ref{J_app_moduli}, \ref{dist_cep_comp}, and Figures \ref{COMPARISON_MOD}, \ref{COMPARISON}. In Figure \ref{COMPARISON_MOD} we compare the apparent distance moduli of our sample of galaxies derived from the Cepheid Leavitt law, to those derived from the JAGB magnitudes in this paper. Figure \ref{COMPARISON} compares the reddening-corrected distances as determined from the Cepheids and carbon stars. It is appreciated that the agreement between the Cepheid and JAGB distances is excellent.
For only one out of eight galaxies, NGC 247, there is a significant disagreement between the distance values derived from the two methods. If we reject the NGC 247, the residuals obtained from subtracting the distance moduli resulting from the two methods yield a very small rms scatter of only $0.06$ mag, and the mean deviation between the distances from the two methods is only $0.01$ mag. Including NGC 247, the rms scatter increases to $0.09$ mag, and the mean deviation is now $-0.02$ mag, still a very small value. The SMC whose eclipsing binary distance \citep{SMC-DEB} is also plotted in Fig. \ref{COMPARISON} plays a control role here and is not included in these statistics. We also note that the JAGB method as calibrated in the LMC produces an excellent agreement between the JAGB distance to the SMC and its distance obtained from detached eclipsing binaries by \citet{SMC-DEB}. We further note that the observed dispersion of the residuals is caused not only by the errors of our JAGB distance determinations described in this work, but by the uncertainties on the previous distance determinations using the Leavitt law as well.

\section{Discussion}

The systematic uncertainties of our distance determinations include the total uncertainty on the LMC distance determination ($0.03$ mag), the uncertainty on the mean JAGB magnitude for the LMC ($0.01$ mag), and uncertainties in our photometric zeropoints for the different galaxies (up to around $0.04$ mag). In total, we estimate the systematic uncertainties on our distances derived from these components to be up to around $0.05$ mag. In our approach, we eliminate many possible sources of systematic errors on relative distances by comparing Cepheid distances and JAGB distances obtained using the same photometry.

We considered also a possible impact of the crowding effect on our determinations by searching for differences in $<J_0>$ obtained for photometric fields of a given galaxy with different densities of sources (with denser fields being closer to the center of the galaxy).
Such tests in the cases of NGC 247 and NGC 55, and NGC 7793 resulted in coherent distance determinations given the photometric zeropoint uncertainty for the $J-$band.
\citet{Bresolin} found, from a comparison of the photometry of Cepheids from Hubble Space Telescope images to their ground-based photometry presented in \citet{N300-CEP}, that the effect of crowding in NGC 300 on the Cepheid distance is limited to less than $0.04$ mag (which is a reasonable assumption given that NGC 55 and NGC 7793 are at distances which are comparable to the distance of NGC 300).
If such an effect occurs in the case of JAGB stars, and its magnitude is similar, it is smaller than the statistical spread of the differences between the JAGB and Cepheid distances we have measured for our sample of galaxies.

Virtually all stars from our samples are variable (\citealt{Weinberg} - based on the LMC population). Their variability is one of the components of the $\sigma$ parameter. \citet{MADORE} estimate their mean J-mag amplitude as $0.7$ mag which corresponds to an equivalent spread of $0.22$ mag. However, the work of \citet{Whitelock}, who investigate AGB stars in NGC 6822, suggests that most stars that constitute our samples are not Mira variables and have low amplitudes.\

Our method yields a precision (estimated from bootstrapping) of almost $0.1$ mag for some samples that have sizes of the order of one hundred stars (WLM is a perfect example). Larger samples allow for a better-defined fit, and they also provide better averaging of phases of variable stars for our statistics (even though their variability is still contained in the $\sigma$ parameter). Small samples are more susceptible to the random density fluctuations of carbon stars in CMDs and possible contamination with foreground stars and background galaxies. Contamination coming from background galaxies has the potential to affect particularly small and distant samples. While our method provides modeling of contamination and distinction of it from the Gaussian term which represents the population of carbon stars, large enough samples are needed to obtain a clear separation of the two components. It is also represented by greater uncertainties for smaller samples as estimated by bootstrapping.

\citep{RIPOCHE} report a relatively large difference between the expected absolute $J-$ magnitudes of carbon stars in the LMC and the SMC, while our values of the LMC and the SMC absolute JAGB magnitudes agree with each other even within the small statistical errors of their determinations. \citeauthor{RIPOCHE} find the difference of the median $J-$band absolute magnitude of the carbon-star luminosity functions for the LMC and the SMC to be $(0.124 \pm 0.016)$ mag, explaining the fainter absolute magnitude in the SMC as due to the lower metallicity of the galaxy which implies fewer dredge-up episodes responsible for converting O-rich stars into C-rich stars given the smaller initial amount of oxygen in the envelope (after \citealt{RicherWesterlund}, \citealt{MGB}, \citealt{Rossi}, \citealt{Mouhcine}, \citealt{M2017}). However, \citet{MADORE} find the difference between the mean of the $J-$ band luminosity functions of carbon stars for the LMC and the SMC to be only $(0.04 \pm 0.02)$ mag. Also, \citet{FREEDMAN} estimate the metallicity impact on the absolute magnitude of JAGB stars to be very small, $(-0.03 \pm 0.05)$ mag/dex, and they note that the effect does not seem to be significant. We note here that all three definitions of the common value of the luminosity function of JAGB stars are different, with our method relying on a parameter of the fitted profile while \citeauthor{MADORE}, \citeauthor{FREEDMAN}, \citeauthor{LEE} use the mean, and \citeauthor{RIPOCHE} as well as \citeauthor{PARADA} rely on the median of the luminosity function, and use selection boxes defined differently - especially along the $J-$ mag axis.

\citet{PARADA} fit the luminosity function as well, adopting a modified Lorentzian distribution. The purpose of this apporach is to estimate the skewness of the luminosity function and thus decide whether a galaxy is more 'SMC-like' or 'LMC-like' in order to choose the proper calibration of the method for a given galaxy. They note that the median of the fitted function does not correlate with any other parameter, but the mode anticorrelates with skewness (bootstrap simulations).
They report that both star formation history and metallicity contribute to the shape, or more specifically skewness, of the luminosity function, but these two components are hard to distinguish.

\begin{figure}[h!]
\plotone{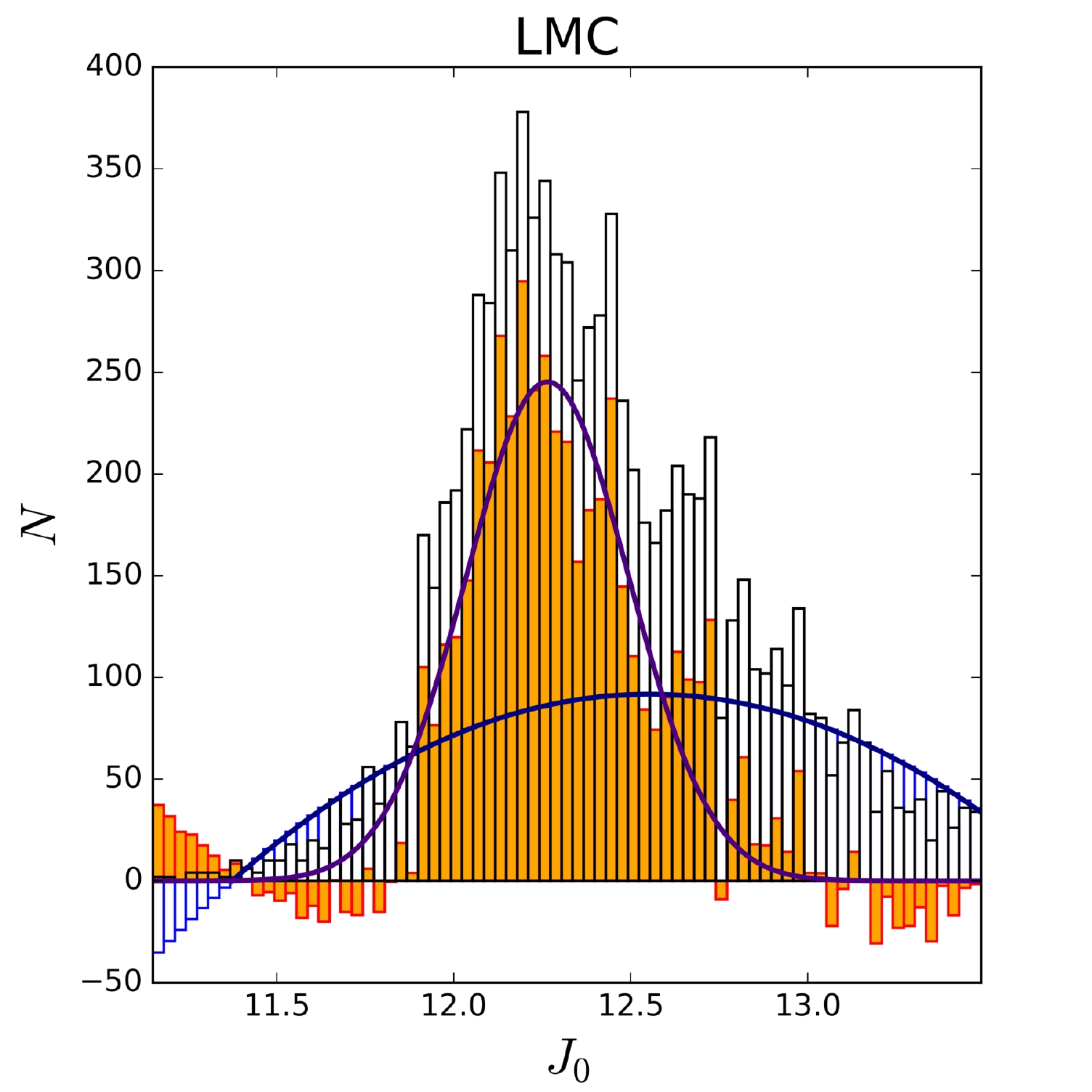}
\caption{Components of the fitted function (Gaussian and quadratic) with the residual distribution (in orange) resulting from the subtraction of the quadratic component of the fit from the apparent distribution of luminosities, for the LMC. The original apparent distribution of luminosities is shown using black, empty bins in the background.}
\label{clean_LMC}
\end{figure}

\begin{figure}[h]
\plotone{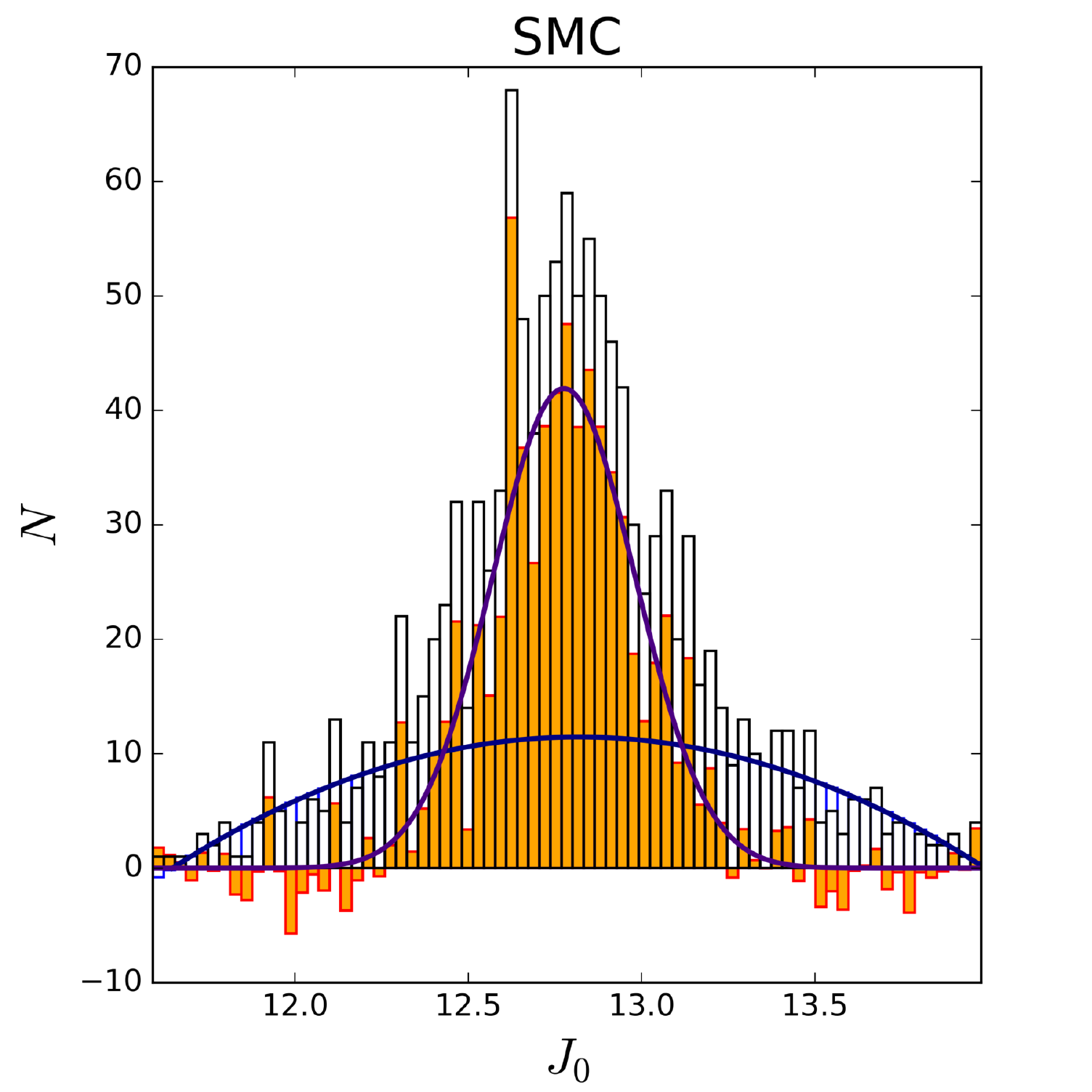}
\caption{As Figure \ref{clean_LMC}, for the SMC.}
\label{clean_SMC}
\end{figure}

\begin{figure}[h!]
\plotone{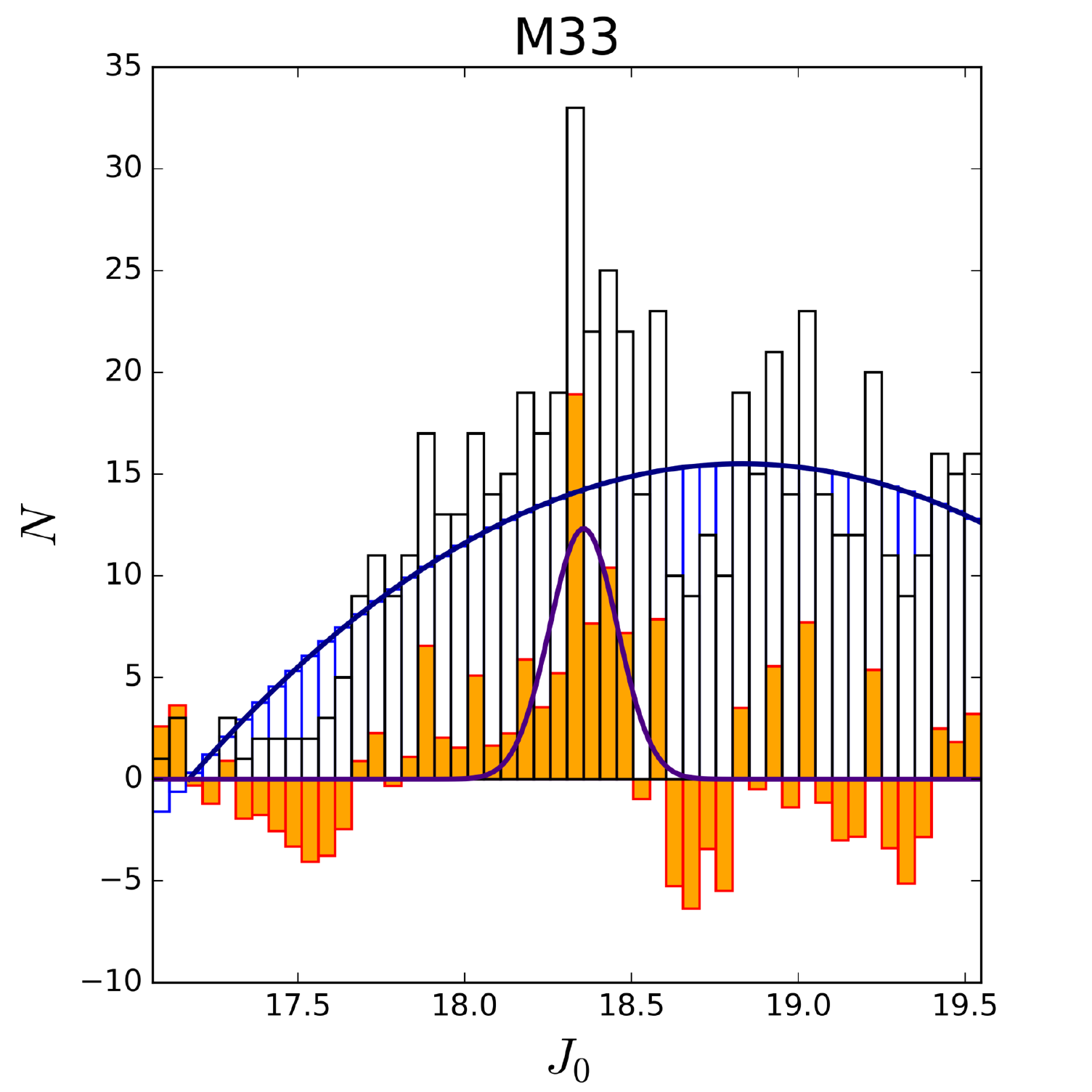}
\caption{As Figure \ref{clean_LMC} and \ref{clean_SMC}, for M33.}
\label{clean_M33}
\end{figure}

In our approach, the skewness of a distribution is contained in the quadratic component of the fitted profile which also takes into account the contamination of a sample. Figures \ref{clean_LMC}, \ref{clean_SMC}, and \ref{clean_M33} show examples of our approach in the case of the LMC, SMC, and M33 respectively. Shifts between the centers of the Gaussian and the quadratic components allow for a rough skewness modelling of an apparent distribution of luminosities. While the residuals resulting from subtraction of the quadratic component of the fit from the observed distribution do not follow perfectly a normal distribution, the fitted function allows for a much more accurate fit than a simple Gaussian. In the case of the SMC, which displays a symmetrical distribution of JAGB luminosities, with small skewness, we see that the quadratic component reproduces the wings of the distribution quite well. A larger skewness as in the case of the LMC or M33 leads to a worse reproduction of the Gaussian distribution but this approach is still useful for the determination of $<J_0>$ - the typical value of the $J-$ magnitude of JAGB stars.


\begin{figure}[h]
\plotone{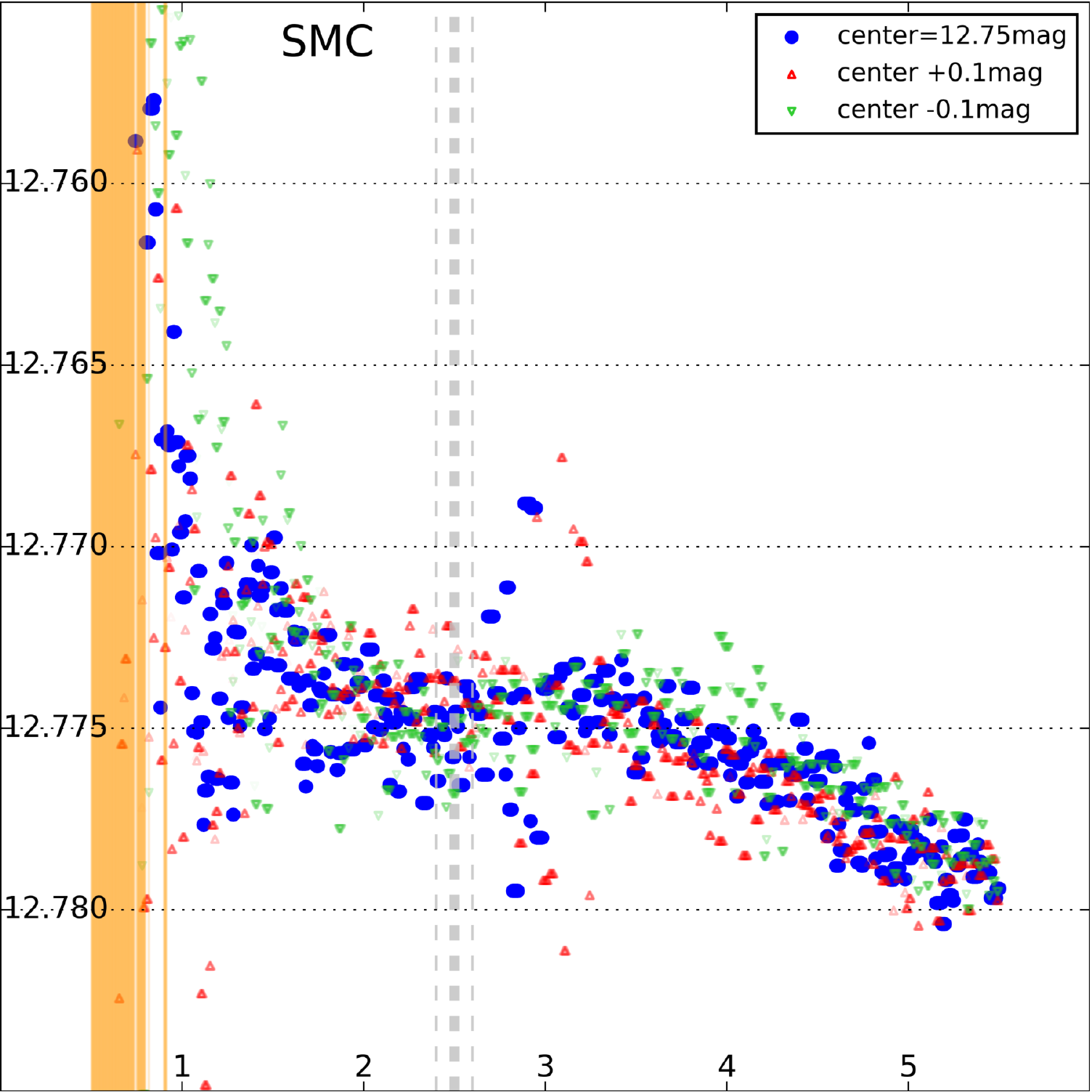}
\caption{Dependence of the determinations of $<J_0>$ using profile fits on selection-box size (x axis, in mag), and position (three different box centers marked with different shapes), for the SMC. The box size of $2.5$ mag chosen for our work (marked by a silver dashed line with a $\pm 0.1$ mag interval denoted with thinner lines) gives a stable result ($<J_0>$, in mag) that does not change significantly with a minor change of the size, and gives universally stable results for the sample of the galaxies we analyzed. Orange areas correspond to selection boxes for which the fit did not converge.}
\label{res_distr_SMC}
\end{figure}

\begin{figure}[h!]
\plotone{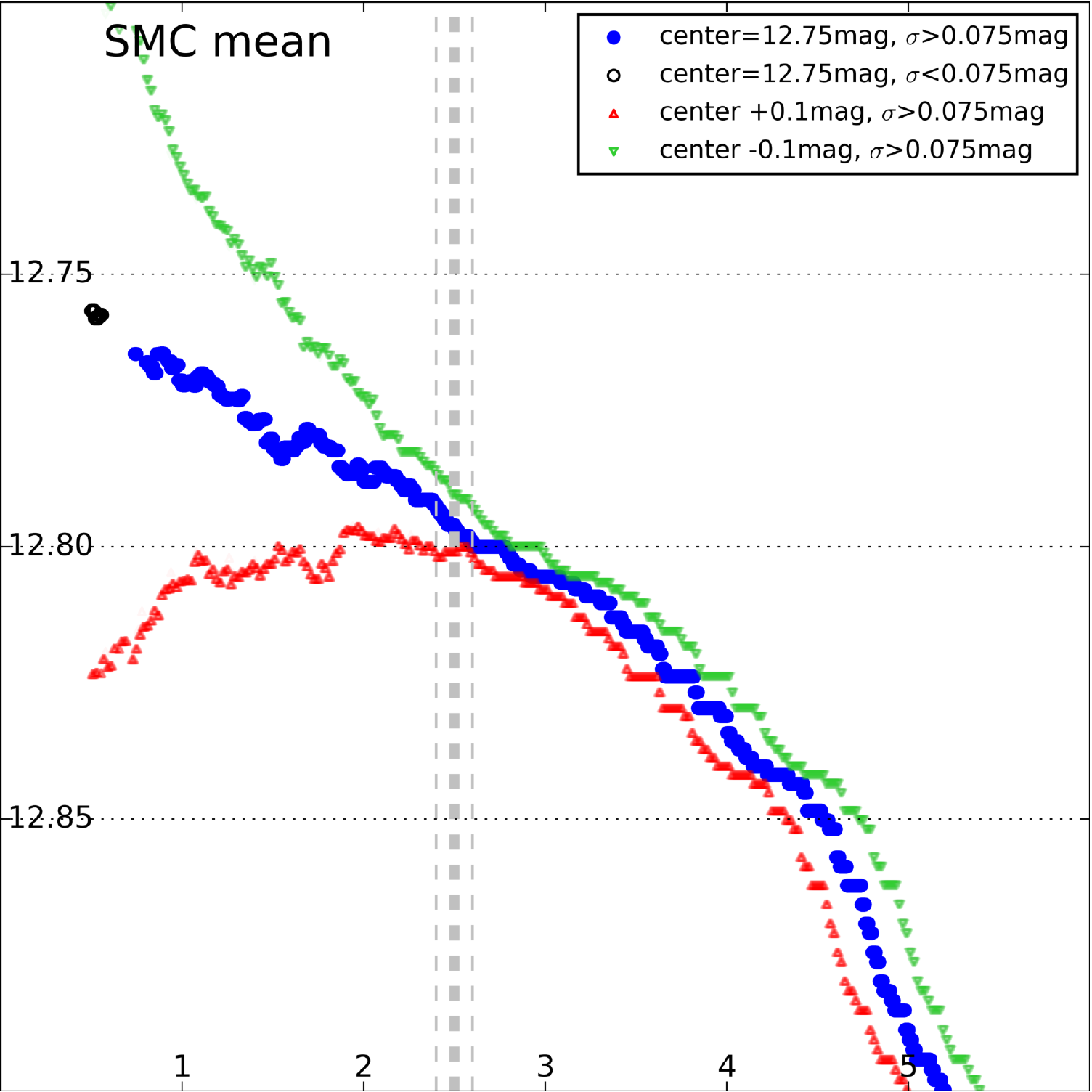}
\caption{As in Figure \ref{res_distr_SMC}, but with determinations obtained using the mean of the $J_0$ magnitudes.}
\label{res_distr_SMCmean}
\end{figure}

\begin{figure}[h!]
\plotone{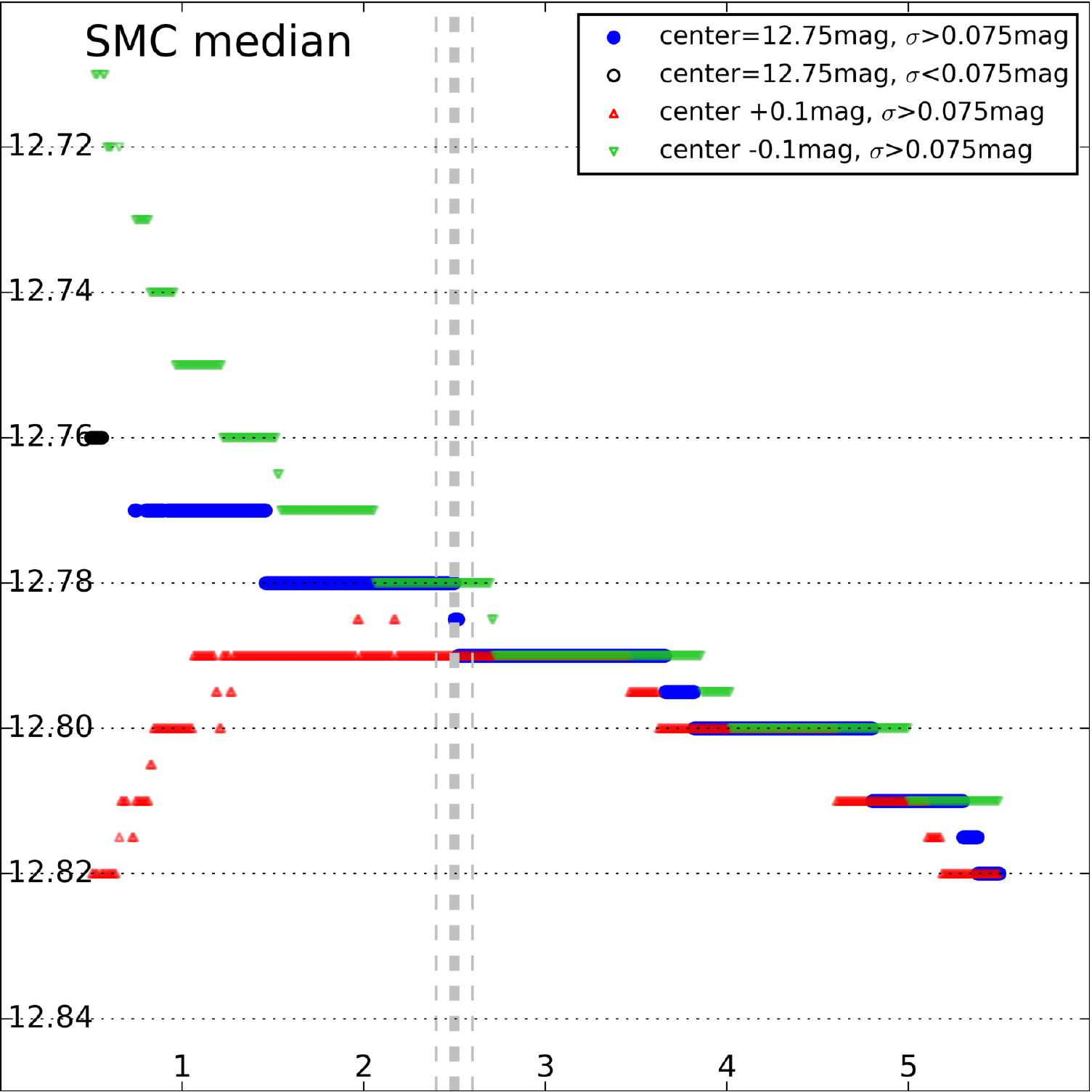}
\caption{As in Figure \ref{res_distr_SMC}, but with determinations obtained using the median of the $J_0$ magnitudes.}
\label{res_distr_SMCmedian}
\end{figure}

\begin{figure}[h!]
\plotone{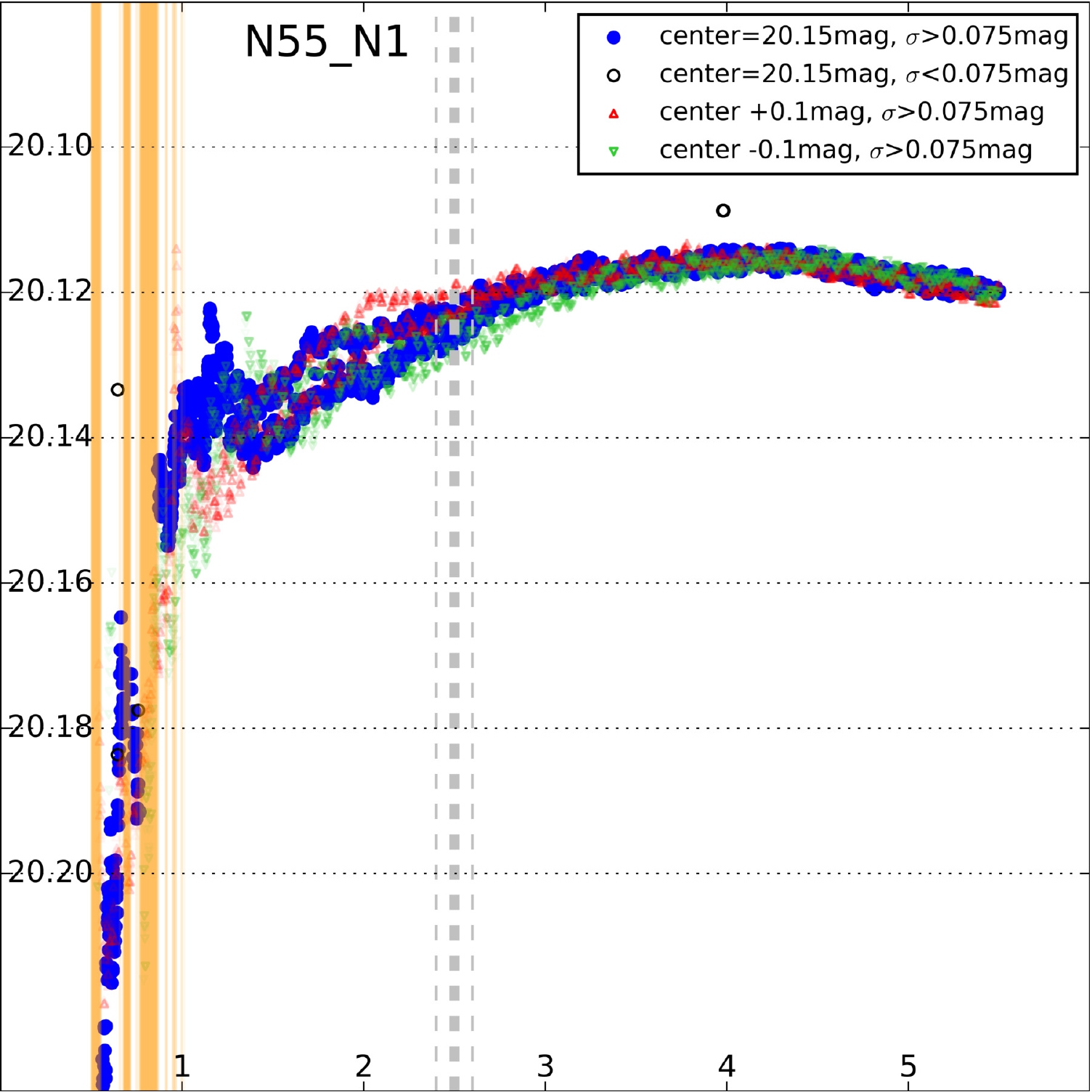}
\caption{Dependence of the obtained $<J_0>$ values on box size and center position for NGC 55. The used $2.5$ mag box width is marked with the thick silver line, and $\pm 0.1$ mag box width variations are marked with thinner lines.}
\label{res_distr_N55_1}
\end{figure}

We have also analyzed the dependence of our results on the size of the sample-selection boxes and the location of their centers, for each galaxy. 
Figures \ref{res_distr_SMC} to \ref{res_distr_SMCmedian} depict these dependences for three different box centers, for the representative case of the SMC. Figure \ref{res_distr_N55_1} shows also results of profile fits for one example outside of the Magellanic Clouds - NGC 55. The relations visible in these figures give an insight into the stability and quality of the $<J_0>$ determinations, and the corresponding distance moduli. 
Orange areas in Fig. \ref{res_distr_SMC}, \ref{res_distr_N55_1} show regions where the fit did not converge. Open dots designate determinations with the $\sigma$ of the Gaussian component of the fitted profile smaller than $0.75$ mag. Small $\sigma$ values appeared usually for fits that detected random fluctuations of density on the CMD and, while in some cases these fits might have been correct (see WLM), we rejected them. In these figures, the blue points indicate the solutions obtained with the finally selected centers of the boxes. Red triangles correspond to box centers $0.1$ mag larger than the value used for the final determination. Green triangles correspond to box centers $0.1$ mag smaller, respectively.

Figures \ref{res_distr_SMC} to \ref{res_distr_SMCmedian} compare the stability of the results obtained from three different statistics - the $m$ coefficient of the fitted function, mean, and median, in the case of the SMC. We clearly see that in the case of median and mean values, the determinations are more sensitive to changes in box sizes and, what's more worrysome, changes in the adopted centers of the sample-selection box. This is especially apparent for small sizes of the sample-selection boxes. While we may fix the same box sizes for all samples, the choice of their centers is independent and arbitrary in the case of each and every sample. We have decided to keep one selection box width (along the $J-$ axis) of $2.5$ mag which gives, importantly, a stable result in the case of our calibrating galaxy, the LMC. Moreover, this choice leads also to stable results for the rest of our sample of galaxies. Our method is also stable with respect to changes in the centers of the sample-selection boxes. 

Finally, we tested our results against different binnings of the data. Most of our determinations (with the WLM being the only exception) were carried out using 1-3 bins per $0.1$ mag. The differences between the JAGB magnitudes for different binnings were usually of the order of $0.01$ mag or even smaller.

Summarizing, the method used in this work yields satisfactory and promising results and should definitely undergo further development and tests - especially arbitrary choices that influence the JAGB magnitude determinations should be limited, and better justified. Also, another form of the fitted luminosity function may be applied in the future. Out of the nine JAGB-based distance determinations for the galaxies scrutinized in this paper, only one galaxy yields a distance result that is significantly at odds with the distance obtained from Cepheids and the same photometric data. The only discrepant JAGB distance, in the case of NGC 247, is currently not understood given the large sample of JAGB stars and the well-fitted luminosity function. The discrepancy is also apparent for the apparent values of distance moduli. Further work is required to resolve this mystery. 

JAGB expected magnitude determinations and diagnostic plots for all analyzed galaxies are available as an appendix to this paper on the following website: \href{https://users.camk.edu.pl/bzgirski/jagb.html}{https://users.camk.edu.pl/bzgirski/jagb.html}.
\\
\\

We thank the anonymous referee for their valuable comments and suggestions.
The research leading to these results has received funding from the European Research Council (ERC) under the European Union’s Horizon 2020 research and innovation program (grant agreement No 695099) and from the National Science Center, Poland grants MAESTRO UMO-2017/26/A/ST9/00446. We acknowledge support from the DIR/WK/2018/09 grant of the Polish Ministry of Science and Higher Education.
We also gratefully acknowledge financial support for this work from the BASAL Centro de Astrofisica y Tecnologias Afines BASAL-CATA (AFB-170002).
RPK acknowledges support by the Munich Excellence Cluster Origins funded by the Deutsche Forschungsgemeinschaft (DFG, German Research Foundation) under Germany's Excellence Strategy EXC-2094 390783311.

\begin{deluxetable}{ccccc}
\tablewidth{0pt}
\tablecaption{Summary of the expected values of the $J-$magnitude of the J-region Asymptotic Giant Branch (JAGB) stars for all analyzed galaxies.}
\tablehead{
\colhead{} & \colhead{$<J_0>_{JAGB}$} & \colhead{$\delta_{JAGB}$} & \colhead{$E(B-V)$} & \colhead{}
\\
\colhead{Galaxy ID} & \colhead{(mag)} & \colhead{(mag)} & \colhead{(mag)} & \colhead{Reddening}
}
\startdata
LMC & 12.265 & 0.010 & 0.139 & \citet{MC-EXT}\\
SMC & 12.776 & 0.012 & 0.089 & \citet{MC-EXT}\\
NGC 6822 & 17.031 & 0.037 & 0.356 &\citet{N6822-CEP}\\
M33 & 18.356 & 0.064 & 0.19 & \citet{M33-CEP}\\
WLM & 18.742 & 0.080 & 0.082 & \citet{WLM-CEP}\\
NGC 3109 & 19.305 & 0.046 & 0.087 & \citet{N3109-CEP}\\
NGC 55 & 20.145 & 0.019 & 0.127 & \citet{N55-CEP}\\
NGC 300 & 20.257 & 0.063 & 0.096 & \citet{N300-CEP}\\
NGC 247 & 21.194 & 0.023 & 0.18 & \citet{N247-CEP}\\
NGC 7793 & 21.489 & 0.027 & 0.08 & \citet{N7793-CEP}\\
\enddata
\label{j0}
\tablecomments{The table contains the expected unreddened magnitudes obtained with the profile fit, and their corresponding statistical uncertainties as obtained from bootstrapping - $<J_0>_{JAGB}$ and $\delta_{JAGB}$, respectively (in mag). The table also includes $E(B-V)$ selective extinctions used to deredden the data for each galaxy and were taken from our previous papers which reported multi-band Cepheid distances. The dereddening for the Magellanic Clouds was done using the \citet{MC-EXT} reddening maps using grid values of $E(B-V)$ - we included mean of the values corresponding to the used fields of \citet{IRSF-CAT} IRSF photometric maps in the table.}
\end{deluxetable}

\begin{deluxetable}{cccccc}
\tablewidth{0pt}
\tablecaption{Comparison of the apparent distance moduli derived in the $J-$band from the JAGB method, and from Cepheids.}
\tablehead{
\colhead{} & \colhead{$\mu_{JAGB}(J)$} & \colhead{$\mu_{CEP}(J)$} & \colhead{$\delta_{JAGB}$} & \colhead{$\delta_{CEP}(J)$} & \colhead{}
\\
\colhead{Galaxy ID} & \colhead{(mag)} & \colhead{(mag)} & \colhead{(mag)} & \colhead{(mag)} & \colhead{Cepheid distance}
}
\startdata
NGC 6822 & 23.545 & 23.563 & 0.037 & 0.025 & \citet{N6822-CEP} \\
M33 & 24.729 & 24.707 & 0.064 & 0.045 & \citet{M33-CEP} \\
WLM & 25.024 & 24.971 & 0.08 & 0.055 & \citet{WLM-CEP} \\
NGC 3109 & 25.591 & 25.627 & 0.046 & 0.028 & \citet{N3109-CEP} \\
NGC 55 & 26.465 & 26.536 & 0.019 & 0.048 & \citet{N55-CEP} \\
NGC 300 & 26.55 & 26.429 & 0.063 & 0.048 & \citet{N300-CEP} \\
NGC 247 & 27.559 & 27.776 & 0.023 & 0.039 & \citet{N247-CEP} \\
NGC 7793 & 27.769 & 27.71 & 0.027 & 0.04 & \citet{N7793-CEP} \\
\enddata
\label{J_app_moduli}
\tablecomments{Comparison of apparent, observed distance moduli in the $J-$band for the case of the method used in this work, $\mu_{JAGB}(J)$, and the Cepheid apparent distance moduli $\mu_{CEP}(J)$ as reported in the original papers, corrected for \citet{LMC-DEB} LMC distance of $18.477$ mag. The uncertainties $\delta_{CEP}(J)$ on the intersections of period-luminosity relations for Cepheids in the $J-$band given in the original papers are adopted as uncertainties on the corresponding apparent moduli resulting from Cepheids. Statistical uncertainties of JAGB moduli are the same as adopted in the case of $<J_0>$ and unreddened distances - $\delta_{JAGB}$.}
\end{deluxetable}

\begin{deluxetable}{ccccccc}
\tablewidth{0pt}
\tablecaption{Summary of distance moduli obtained in this work, and their comparison to Cepheid distances.}
\tablehead{
\colhead{}  & \colhead{$\mu_{0, JAGB}$} & \colhead{$\mu_{0, CEP}$} & \colhead{$\mu_{0, EB}$} & \colhead{$\delta_{JAGB}$} & \colhead{$\delta_{CEP}$} & \colhead{$\delta_{EB}$}
\\
\colhead{Galaxy ID}  & \colhead{(mag)} & \colhead{(mag)} & \colhead{(mag)} & \colhead{(mag)} & \colhead{(mag)} & \colhead{(mag)}
}
\startdata
SMC & 18.988 & (...) & 18.977 & 0.012 & (...) & 0.016  \\
NGC 6822  & 23.243 & 23.289 & (...) & 0.037 & 0.021 & (...) \\
M33  & 24.568 & 24.60 & (...) & 0.064 & 0.03 & (...) \\
WLM  & 24.954 & 24.901 & (...) & 0.08 & 0.042 & (...) \\
NGC 3109 & 25.517 & 25.548 & (...) & 0.046 & 0.024 & (...)\\
NGC 55  & 26.357 & 26.411 & (...) & 0.019 & 0.037 & (...) \\
NGC 300  & 26.469 & 26.344 & (...) & 0.063 & 0.04 & (...) \\
NGC 247  & 27.406 & 27.621 & (...) & 0.023 & 0.036 & (...)\\
NGC 7793  & 27.701 & 27.65 & (...) & 0.027 & 0.04 & (...)\\
\enddata
\label{dist_cep_comp}
\tablecomments{Distance moduli resulting from JAGB determinations using the profile fit, and determination using the Leavitt law for Cepheids, and their corresponding errors are denoted by $\mu_{0, JAGB}$, $\mu_{0, CEP}$, $\delta_{JAGB}$, and $\delta_{CEP}$, respectively (in mag). The table also includes the SMC distance $\mu_{EB} \pm \delta_{EB} $ from eclipsing binaries \citep{SMC-DEB}. All uncertainties are statistical.}
\end{deluxetable}
\clearpage
\newpage
\end{document}